\documentclass[a4paper,fleqn]{cas-sc}

\usepackage[numbers]{natbib}
\def\tsc#1{\csdef{#1}{\textsc{\lowercase{#1}}\xspace}}
\tsc{WGM}
\tsc{QE}

\begin{document}
\let\WriteBookmarks\relax
\def\floatpagepagefraction{1}
\def\textpagefraction{.001}

\shorttitle{Research progress on QNNs and QML}    

\shortauthors{Y. Sun, et al}  

\title [mode = title]{Research progress on quantum neural networks and quantum machine learning}  



%

\author{Yifan Sun}[orcid=0000-0003-4935-630X]
\ead{yfsun@bit.edu.cn}
\credit{Writing the main content of the paper}

\author{Boyuan Sun}
\ead{BoYuan_Sun@outlook.com}
\credit{Drafting sections of the paper, collecting references}

\author{Jiameng Tian}
\ead{tjm0x7d3@vip.qq.com}
\credit{Drafting sections of the paper, collecting references}

\author{Xiangdong Zhang}
\ead{zhangxd@bit.edu.cn}
\credit{Initialization of the paper}
\cormark[1]
\cortext[1]{Corresponding author}

\affiliation{organization={Key Laboratory of Advanced Optoelectronic Quantum Architecture and Measurements of Ministry of Education, Beijing Key Laboratory of Nanophotonics and Ultrafine Optoelectronic Systems, School of Physics, Beijing Institute of Technology},
            city={Beijing},
            postcode={100081}, 
            country={China}}

\begin{abstract}
Machine learning holds fundamental computational significance due to the increasing demand for efficient solutions to complex tasks in data analysis, pattern recognition, and optimization, which are essential for addressing the multifaceted challenges of modern society. As the volume of data proliferates at an unprecedented rate, the need for more powerful machine learning strategies becomes increasingly evident. Quantum neural networks (QNNs) represent an emerging and transformative research field that seeks to harness the unique principles of quantum mechanics to enhance the capabilities of machine learning algorithms. This survey examines various QNN approaches, including fully connected QNNs, quantum convolutional neural networks, equivariant QNNs, quantum Hopfield networks, quantum Boltzmann machines, quantum reservoir computing, and composite networks for quantum reinforcement learning, quantum generative learning, and quantum transfer learning. We summarize the relevant investigations on their performance, including learning accuracy, training time, and resource requirements, etc. Each QNN type has unique strengths and weaknesses, offering diverse solutions for different applications.
\end{abstract}



\maketitle
\tableofcontents
\section{Introduction}\label{Sec:Intro}

Machine learning is of great importance for the data science of today, and classical neural networks (NNs) have turned out to be indispensable tools for handling such tasks \cite{Kussul2010,Lecun2015}. Inspired by the architecture of the human brain, these networks consist of interconnected nodes organized into layers, enabling them to learn intricate patterns from large datasets. Their versatility has led to widespread adoption across diverse applications, including image and speech recognition, natural language processing, and predictive analytics. Classical NNs not only have enhanced the efficiency and accuracy of these tasks but also have revolutionized industries and daily life by empowering machines to learn from experience and make data-driven decisions. However, as data volumes continue to grow exponentially, classical NNs face significant limitations in computational efficiency and scalability. This has activated the search for more advanced approaches to meet the increasing demands of modern data processing. 

In the pursuit of highly efficient models for learning patterns of data, quantum neural networks (QNNs) have emerged as one of the most promising candidates. These networks aim to transcend the limitations of classical NNs by leveraging the principles of quantum mechanics \cite{Biamonte2017,Beer2020}. By utilizing quantum bits (qubits) and exploiting quantum phenomena such as superposition and entanglement, QNNs can potentially process vast amounts of data more efficiently than their classical counterparts. This capability enables faster training times and improved accuracy in handling complex tasks, making them well-suited for high-dimensional data and intricate optimization problems. Such advantages are particularly critical for applications ranging from drug discovery to financial modeling, where the ability to process and analyze large datasets quickly and accurately is paramount. 

\begin{figure}\centering
\includegraphics[width=.9\textwidth]{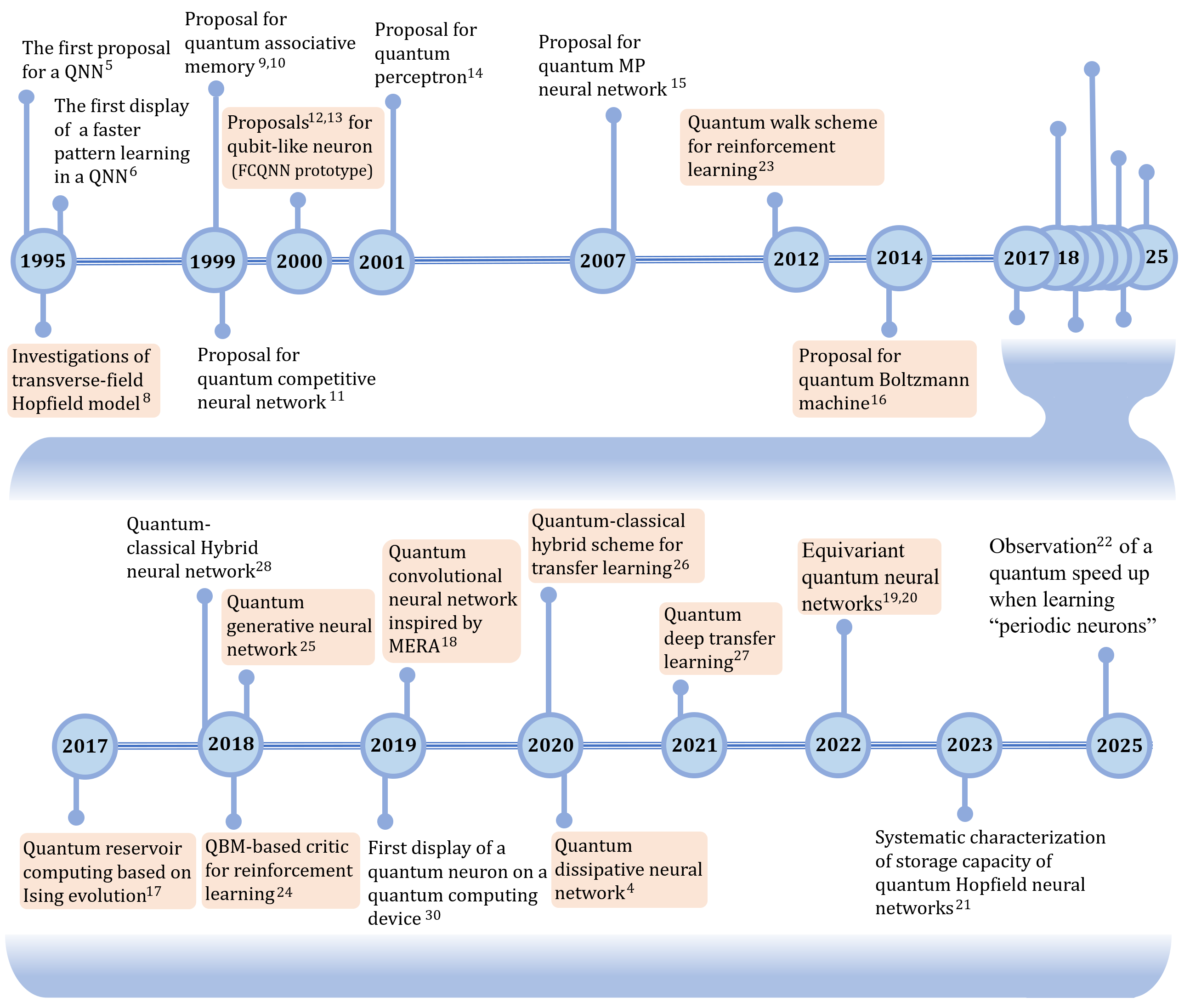}
\caption{Timeline highlighting major advancements several basic structures of QNNs and quantum machine learning. Those ones highly relevant to this review are marked out by colored background.}\label{FIG:1}
\end{figure}

A timeline highlighting major advancements in QNNs is shown in Fig. \ref{FIG:1}. Historically, the original motivation for investigating QNNs is to explain brain functions through quantum principles. The works focusing on such an aim were published from the mid-1990s to 2007, known as the first phase of the relevant researches. Particularly, the idea of QNN was first introduced in 1995 by combining artificial neural networks with quantum computation, the significant contributions of which come from Kak \cite{Kak1995}, Menneer and Narayanan \cite{Menneer1995}, demonstrating faster pattern learning in QNNs compared to classical NNs. During the same period, a quantum version of the Hopfield NN is proposed \cite{Ma1993,Ma1995}, starting the relevant investigations with the examples of transverse-field Ising model. Other notable contributions during this period include the development of quantum associative memory \cite{Ventura1999,Peruš2000}, quantum competitive NN \cite{Ventura1999b}, QNN based on qubit-like neuron \cite{Matsui2000,Kouda2005}, quantum perceptron models \cite{Altaisky2001}, quantum McCulloch-Pitts NNs \cite{Zhou2007}, etc. Thereafter, together with the development of quantum control technologies, the research on QNNs has entered the second phase, which began around 2007 and continues to the present. In such a phase, a practically workable form of QNNs is highly concerned, focusing on the performance of data processing rather than a closely mimic of brain functions. This period has seen a surge in interest and practical advancements, and numerous novel types of QNNs and pathways to their implementation are reported. Those achievements can be categorized by three exploration directions. In the first direction, researchers focus on proposing different QNNs and broadening the applications to different types of datasets. For example, the proposals of quantum Boltzmann machines (QBMs) \cite{Wiebe2014}, quantum reservoir computing (QRC) \cite{Fujii2017}, quantum convolutional NNs (QCNNs) \cite{Cong2019}, quantum dissipative NNs (QDNNs) \cite{Beer2020}, equivariant QNNs (EQNNs) \cite{Skolik2023,West2023}, and other works \cite{Bödeker2023,Lewis2025}, are presented following the line. In the second direction, researchers try to define complete learning formalism based on the properties of QNNs, or even the quantum evolutions. Typical examples include investigations on quantum reinforcement learning (QRL) \cite{Briegel2012,Crawford2018}, quantum generative learning (QGL) \cite{Lloyd2018}, quantum transfer learning (QTL) \cite{Gokhale2020,Wang2021}, etc. In the third direction, researchers pay attention to the development of learning strategies based on more practically available systems. Along this line, important achievements include hybrid quantum-classical learning schemes \cite{Gardas2018,Broughton2020}, implementing neurons on a quantum processor \cite{Tacchino2019}, etc.

Here, we provide a broad overview that focuses on the first and second directions, categorizing QNN models by their typical characteristics, and reviewing their applications for different kinds of machine learning. When discussing practical applications, we involve the researches on the third direction a bit, but does not review them in details. There have been reviews of QNNs from the perspective of timelines \cite{Chakraborty2020}, concepts \cite{Alchieri2021}, models \cite{Zhao2021,Mangini2021}, and publications \cite{Valdez2023}. In contrast, this review offers a systematic and methodological streamline of the field, incorporating seminal contributions in recent years.

\begin{figure}\centering
\includegraphics[width=.8\textwidth]{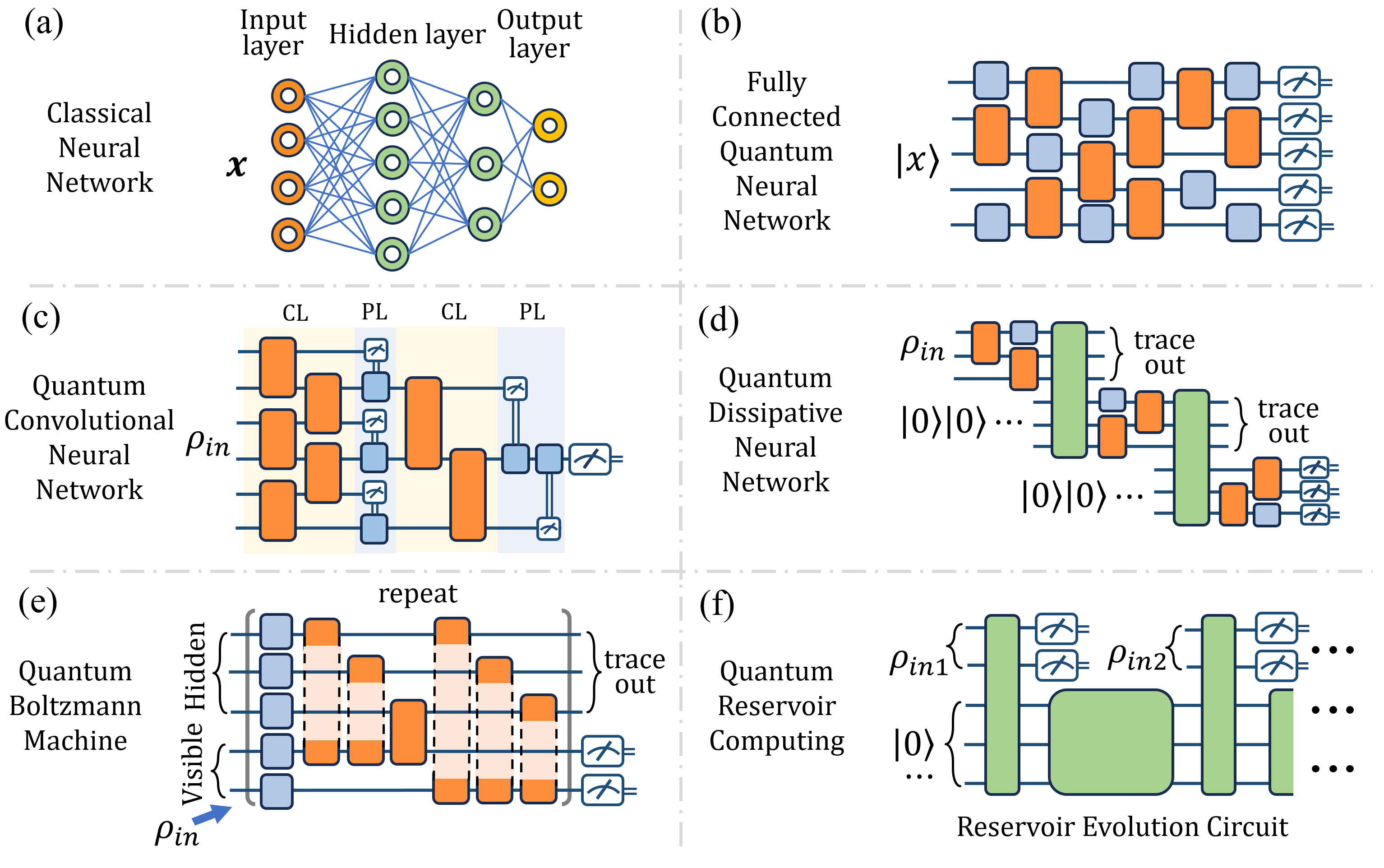}
\caption{(a) A sketch of a Classical NN, with input data vector $x$. (b) The FCQNN, with input data carried by quantum state $|x\rangle$. (c) The QCNNs. $\rho_{in}$ denotes the density matrix of the input. (d) The QDNNs. (e) The QBMs. Several qubits are used as the hidden ones, and others are used as the visible ones. (f) The QRCs. The blue blocks, orange blocks, and the green blocks represent single qubit gates, two-qubit gates, and multi-qubit gates, respectively.}\label{FIG:2}
\end{figure}

The content of the review is organized as follows. First, we briefly summarize the core of QNNs and the training strategies for them. Due to the special properties of QNNs, the training strategies \cite{Abbas2023,McClean2018,Larocca2025,Crooks2019,Banchi2021,Wierichs2022,Stokes2020,Tao2023,Spall1992,Kandala2017,Lamata2018,Li2024,Nelder1965,Aminpour2024,Bonet-Monroig2023} are not entirely the same as those applied to classical NNs. We provide an instruction on the methods, conceptually explaining the technical lines. The typical structures of QNNs we consider are briefly illustrated in Fig. \ref{FIG:2}. The structure of a classical NN, as the conceptual foundation of QNNs, is given in Fig. \ref{FIG:2}(a). Then, QNNs are categorized based on the types of structural requirements they impose, such as fully connected QNNs (FCQNNs), which have no special requirements (Fig. \ref{FIG:2}(b)), QCNNs, which involve measurement-based rotations (Fig. \ref{FIG:2}(c)), QDNNs, which allow the insertion and removal of qubits (Fig. \ref{FIG:2}(d)), QBMs, which utilize Ising gates (Fig. \ref{FIG:2}(e)), and QRC, which relies on "recurrent" quantum evolutions (Fig. \ref{FIG:2}(f)). It is worth mentioning that the quantum Hopfield networks (QHNs) are structurally similar to QBMs despite their distinct concerns for learning tasks. Therefore, QHNs are not included in Fig. \ref{FIG:2} and will be reviewed with QBMs in the following section. The idea of QDNNs has a close connection with QCNNs, yet they are structurally distinct from QCNNs. Therefore, QDNNs are reviewed together with QCNNs in the following section, although they are illustrated in different panels of Fig. \ref{FIG:2}. The requirements on the EQNNs lie in the symmetry of the data, which is rather abstract and can be applied to many kinds of specific QNN structures. Therefore, it is reviewed without being illustrated in Fig. \ref{FIG:2}. Then, we review the composition of the QNNs for advanced types of machine learning, leading to QRL \cite{Dong2008,Paparo2014}, QGL \cite{Lloyd2018}, and QTL \cite{Marias2021,Wang2021}. As a reference, we also briefly summarize the quantum learning strategies without QNNs, which have a significant influence on the development of quantum machine learning. Finally, we present very recent theoretical discussions on QNNs, including those relevant to the barren plateau problem \cite{McClean2018,Bermejo2026}. 

\section{Basics for QNNs}\label{Sec:Basics}
\subsection{General Descriptions}
\label{subsec:GD}
As we have mentioned, the general purpose of proposing QNNs is to model the patterns, logic, or other characteristics of data using a potentially more powerful and universal framework. Under this purpose, the learning process of QNNs shares several similarities with classical learning theory. Fundamentally, like classical NNs, QNNs are a class of functions designed to "fit" the underlying patterns of given data. They are also typically composed of layerwise structures, which have proven to be efficient in nearly all forms of information processing. Such a kind of architectural principle has been widely validated in both classical deep learning and quantum circuit design. Moreover, as in all kinds of fitting, QNN training has either a specific target (in supervised learning) or an abstract target (in reinforcement learning, generative learning, transfer learning, or unsupervised learning). Consequently, one must tune the parameters of a QNN so that the deviation (or bias) of the model output from the target output is minimized. This minimization is achieved through loss functions, alignment measures, or other benchmarking strategies, mirroring the optimization paradigm of classical machine learning while also introducing specific adaptations for quantum.


From the perspective of data flow, what significantly differentiates QNNs from classical NNs is that, in QNNs, data is stored in a special type of physical state, i.e., quantum states. Consequently, its processing is represented by operators acting on the space of those quantum states, i.e., operators on a Hilbert space. It is worth mentioning that the Hilbert space is not a concept uniquely attached to the quantum world. In fact, classical NNs also operate on feature spaces that are Hilbert spaces. Thus, the mere fact of using a Hilbert space is not the key difference. To see this more clearly, we first specify two basic concepts in machine learning: the \textbf{data space} and the \textbf{feature space}. The \textbf{data space} is the original input domain, such as images, sensor readings, or text embeddings. It is typically not required to be a vector space, nor does it need a well-defined inner product or completeness. The \textbf{feature space} is the internal representation space learned by the network. The main goal of machine learning could be formulated as finding a proper operator path for the input within the feature space. Notably, for both classical NNs and QNNs, the feature space is a finite-dimensional Hilbert space.

\begin{figure}\centering
\includegraphics[width=.75\textwidth]{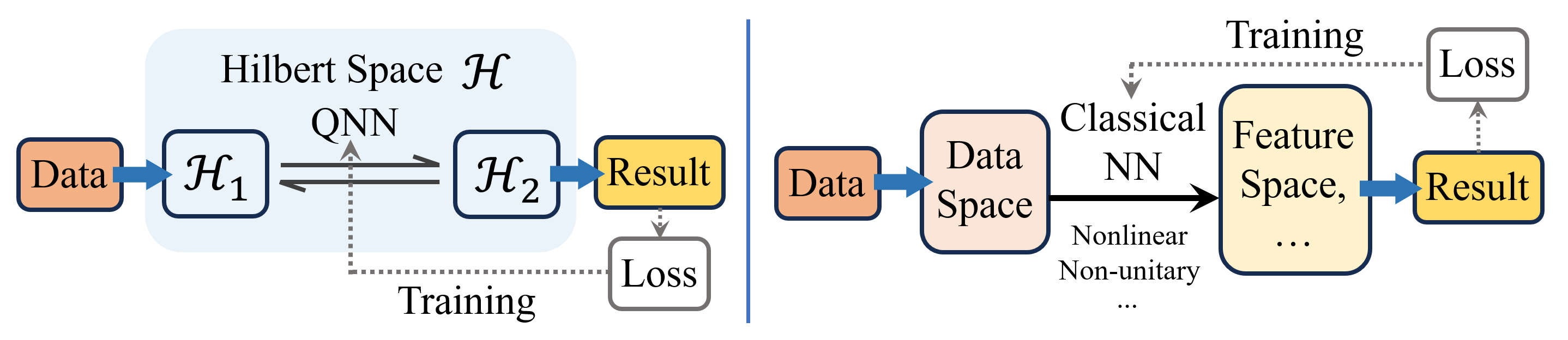}
\caption{The learning process of a QNN vs a classical NN, in a generalized perspective of data flow. By using a QNN (the left panel), the data is encoded by quantum states (in the subspace $\mathcal{H}_1$), and transformed to other states (in the subspace $\mathcal{H}_2$). The training is done by computing the loss function of the output states, and updating the parameters of the QNN. By using a classical NN (the right panel), the data is encoded by the states of digital electrical circuit states
. The size of the state space when data is processed often changes. The training is also done by computing the loss function of the output states, and updating the model parameters.}\label{FIG:3}
\end{figure}

On this foundation, the first difference emerges. A classical NN itself (especially its front layers) acts as an embedding map that progressively transforms raw data from the data space into a high-level feature space. In QNNs, by contrast, the embedding procedure, i.e., encoding classical data into quantum states, is usually simple and fixed, such as via amplitude encoding, angle encoding, or a hardware-efficient ansatz. The core processing then occurs purely within the quantum Hilbert space through unitary operations. This means that in QNNs, the feature space is explicitly a Hilbert space from the very beginning of processing, whereas in classical NNs, the feature space emerges gradually and becomes more complicated through layers that are not constrained by Hilbert space axioms. This fact also leads to the second difference. The data processing enabled by a QNN is mostly unitary, unless measurements on qubits are introduced. The unitary property guarantees both norm preservation and reversibility of the data flow. For classical NNs, the layers can implement arbitrary, often non-invertible functions. Third, classical NNs suffer from gradient explosion and vanishing gradients, which arise from the product of Jacobian matrices during backpropagation. This phenomenon is independent of the completeness of the feature space and is typically addressed via techniques such as gradient clipping, residual connections, or careful initialization. QNNs, on the other hand, face a distinct challenge known as the barren plateau. Due to the structure of random parameterized quantum circuits, the variance of gradients decays exponentially with the number of qubits, making optimization nearly impossible for large-scale QNNs without careful circuit design or classical pre-training. While both problems relate to the flow of gradients through the network, their origins are fundamentally different. See Fig.~\ref{FIG:3} for a comparison of the generalized learning process of the two architectures.

There are often misunderstandings regarding the practical potential of QNNs. As noted, QNNs are proposed for learning from data. However, the vast majority of data in the modern world is encoded by the states of digital electrical circuits. Therefore, to apply QNNs to model real world data patterns, the physical carrier of data must be firstly converted from classical circuits into a system that supports the quantum computing paradigm, then processed, and finally read out and converted back to classical circuits. This round trip conversion typically imposes a substantial overhead that is absent in classical learning. Given this fact, the practical potential of QNNs should be understood in two senses. 

In the first sense, hardware are highly concerned. The device running the QNN would have a significant advantage over that running a classical NN. At present, this remains unclear because no actual quantum computer that can execute the proposed QNNs for practical learning. Nevertheless, if one did possess a quantum computer, how might the potential emerge? For example, if data could be natively stored as quantum states and directly used as training input for a QNN, one could make a fair comparison between QNNs and classical NNs. It shall be done in terms of parameter count, number of blocks, layers, etc., achieving the same learning accuracy. Such a comparison would rest on the assumption that the running time of a quantum computing unit is comparable to that of a classical one. For example, one quantum gate (e.g., a Pauli rotation gate or CNOT) versus one classical gate (e.g., an adder or multiplier), or one quantum module versus one classical electrical circuit, etc. If this assumption does not hold, then the encoding process becomes critical. The time and computational cost of encoding must be bearable relative to the total cost, such that running a QNN still consumes less overall than running a classical NN. In principle, this could also work. However, as far as we know, none of these conditions are met today. Consequently, most results concerning QNNs are actually expectations of how well we might perform once a quantum computer becomes available.

The second sense is approximately the situation we currently find ourselves in. Researchers can consider QNNs as new types of models with additional constraints when learning from data. Notably, most contemporary QNNs are simulated on classical computing devices. From a pure computational perspective, these simulated QNNs function as novel pattern fitting architectures that enrich the available model zoo. As a supporting point, many recent advances in classical machine learning have come from borrowing principles originally developed for quantum systems. Meanwhile, because qubit systems are naturally high‑dimensional, the quantum benchmarks and mathematical structures used in QNN research also aid explorations of modeling high‑dimensional data patterns, such as that encountered in large language models or computer vision. Thus, using a QNN, even a simulated one, is akin to introducing a new class of functions for fitting data. Such an endeavor remains important, because finding a new, potentially more efficient way to express data patterns is always difficult and of great value.

To this point, reviewing the diverse structures of QNNs also inspires novel modeling approaches for the large scale data increasingly collected from around the world. Even if practical quantum advantage remains a long‑term goal, the structural innovations driven by QNN research contribute to a broader research topics, such as data modeling and hybrid algorithms, as well as theoretical insights that benefit both quantum and classical machine learning communities.

\subsection{Training Methodologies}\label{subsec:TM}
As noted, there are both differences and similarities between the training of a QNN. We briefly summarize the algorithmic techniques in this subsection. Like classical machine learning, the main target of training a QNN is also to minimize a cost function $\mathcal{L}(\boldsymbol{\theta})$, and $\boldsymbol{\theta}$ is the parameter vector of the QNN. The update of $\boldsymbol{\theta}$ can also be viewed as finding the steepest path for minimizing $\mathcal{L}(\boldsymbol{\theta})$. However, the training of QNNs is faced with unique challenges, such as difficulties in direct gradient back propagation \cite{Abbas2023}, exponential vanishes of gradients (usually called the barren plateaus problem, and we summarize the topic in the last part) \cite{McClean2018,Larocca2025}, etc. These obstacles have motivated the development of specialized optimization techniques. Broadly, the techniques for achieving so can be categorized into gradient-based methods, which attempt to estimate the gradient despite quantum constraints, and gradient-free methods, which rely on direct cost evaluations.

Firstly, we review the gradient-based methods, which mainly includes finite-difference methods, parameter-shift rule methods, and quantum natural gradient descent methods.

\textbf{Finite-difference (FD)} methods directly estimate gradients via numerical evaluation of the cost function at shifted parameter values. The basic approach is straightforward. For each parameter $\theta_i$ in a QNN, one evaluates the cost function twice. Once with a small positive shift $\theta_i+\epsilon$ and once with a negative shift $\theta_i-\epsilon$ (or a single-sided difference). Then, the gradient of the function $\mathcal{L}$ is approximated as 
\begin{equation}
\frac{\partial\mathcal{L}}{\partial\theta_i}\approx\frac{\mathcal{L}(\theta_i+\epsilon)-\mathcal{L}(\theta_i-\epsilon)}{2\epsilon}.    
\end{equation}
This is the most basic numerical differentiation and can be extended to arbitrary multivariate setting.

Obviously, the methods require no modifications to the quantum circuit structure nor any knowledge of the circuit's internal gate derivatives, so that they are easy to apply. However, these methods suffer from a fundamental limitation. Their time complexity scales linearly with the number of parameters, i.e., $O(p)$ cost evaluations for $p$ parameters. This is not very efficient compared to other methods (such as the ones below). More importantly, on near-term quantum hardware, FD methods are notoriously sensitive to noise. Each evaluation of the cost function accumulates statistical sampling error and device noise. Subtracting two noisy values amplifies the error, especially when the shift $\epsilon$ is chosen too small (leading to catastrophic cancellation) or too large (introducing truncation error). This trade-off makes FD gradient estimation impractical for deep circuits, where parameter counts can reach thousands, or for noisy quantum hardware, where shot noise dominates. 

\textbf{Parameter-shift rule (PSR)} methods that directly evaluate the exact gradient of expectation-based quantum cost functions. Unlike FD methods for numerical gradient approximation, the PSR exploits the analytic structure of quantum expectation values to compute exact gradients. For a parameter $\theta$ of a gate $e^{-i \theta G}$, where $G$ is a Hermitian generator with eigenvalues $\pm r$, the gradient of the expectation value $f(\theta)=\langle\psi(\theta)|O|\psi(\theta)\rangle$ takes a closed form, where $O$ is an operator for benchmarking the loss. Specifically, the PSR requires executing the same circuit twice with shifted parameter values,
\begin{equation}
\frac{\partial f}{\partial\theta}=r\left[f\left(\theta+\frac{\pi}{4r}\right)-f\left(\theta-\frac{\pi}{4r}\right)\right],    
\end{equation}where $f$ denotes the expectation value measured from the circuit output, and $r$ is a constant that depends on the generator $G$. For the common case where the generator has eigenvalues $\pm 1$, e.g., Pauli gates $X,~Y,~Z$, $G$ is usually set to be $X/2$, $Y/2$, or $Z/2$. Then, we have $r=1/2$, and the shift simplifies to $\pi/2$. 

The PSR offers a significant advantage over FD methods. It completely avoids the instability associated with numerical differentiation. There is no need to choose a small $\epsilon$ that trades off truncation error against catastrophic cancellation. Also, the PSR computes the exact gradient using a fixed and deterministic shift. The error of the methods scale up to statistical sampling noise from finite shot counts. This makes PSR substantially more reliable on both simulated and actual quantum hardware. Subsequent developments have extended the original rule to handle a broader class of gates, including those with generators having more than two distinct eigenvalues, which require more than two shifted evaluations, and parameterized multi-qubit gates \cite{Banchi2021,Wierichs2022}.

Unfortunately, the PSR shares a major limitation with FD as well. Its computational cost scales linearly with the number of parameters. For a circuit with $p$ parameters, evaluating the full gradient vector requires $2p$ expectation-value measurements, because there are two shifted circuits per parameter. This linear scaling becomes prohibitive for deep QNNs or large-scale variational quantum algorithms, where $p$ can reach thousands or more. Furthermore, the PSR often requires complex gate decompositions when applied to multi-qubit circuits or gates that do not have a simple Hermitian generator with known eigenvalues. For example, certain hardware-efficient ansatz or custom parameterized gates may not admit a straightforward two-term shift rule, necessitating more elaborate decomposition strategies or hybrid approaches that combine PSR with other gradient estimation techniques.

Despite these challenges, the PSR remains the gold standard for analytic gradient computation in a QNN, striking a balance between accuracy and implementability on near-term quantum devices.

\textbf{Quantum natural gradient descent (QNGD)} methods incorporate the geometry of quantum state space into variational optimization \cite{Stokes2020}. Unlike standard gradient descent, which assumes a Euclidean parameter space and thus follows the steepest descent direction in the parameter norm, QNGD accounts for the fact that the parameterization of a quantum circuit induces a non-Euclidean geometry on the manifold of quantum states. This method captures how infinitesimal changes in parameters affect the underlying quantum state, guiding updates along directions of maximal physical change rather than merely maximal parametric change. The key geometric object is the Fisher information metric, which measures the distinguishability between a pair of quantum states that are close to each other.

Given a cost function $\mathcal{L}(\boldsymbol{\theta})$ and the quantum Fisher information matrix $\mathbf{G}(\boldsymbol{\theta})$, also known as the Fubini-Study metric tensor, the natural gradient update rule is given by
\begin{equation}
\boldsymbol{\theta}_{t+1}=\boldsymbol{\theta}_t-\eta\cdot\mathbf{G}^{-1}(\boldsymbol{\theta}_t)\nabla\mathcal{L}(\boldsymbol{\theta}_t),    
\end{equation}where $\eta$ is the learning rate and $\nabla \mathcal{L}(\boldsymbol{\theta})$ is typically computed using PSR or other analytic gradient methods. Intuitively, the inverse Fisher matrix $\mathbf{G}^{-1}$ rescales the standard Euclidean gradient to account for the curvature in the state space. Hence, the parameter directions that produce large changes in the quantum state are stepped more cautiously, while directions that produce little effect are amplified.

Notably, QNGD exhibits several desirable properties. First, it is invariant under reparameterization. The update trajectory induced by QNGD does not depend on how the circuit parameters are chosen. It depends only on the underlying family of quantum states. This is a crucial advantage over standard gradient descent, which can behave poorly under different but equivalent parameterizations. Second, QNGD can significantly accelerate convergence, especially in regions where the parameter landscape is highly curved or ill-conditioned. Third, under specific conditions, QNGD has been shown to mitigate the barren plateau phenomenon \cite{Tao2023} in certain cases. The reason is that, in those cases, natural gradient can effectively whiten the gradient noise and follows a direction that remains informative even when individual partial derivatives become exponentially small.

However, QNGD faces a major practical bottleneck. The full evaluation of $\mathbf{G}(\boldsymbol{\theta})$ scales quadratically with the number of parameters. For a circuit with $p$ parameters, the quantum Fisher information matrix is a $p \times p$ matrix, requiring $O(p^2)$ measurements or circuit executions to estimate all its entries, or $O(p)$ of them under certain structured approximations. It is generally hard to achieve for large $p$. Moreover, inverting this matrix adds an additional $O(p^3)$ computational cost if done directly, although iterative methods can mitigate this. Consequently, while QNGD is theoretically elegant and powerful, its lack of efficiency in practice limits its application to small-scale variational circuits (typically $p\lesssim 10^2$). Recent research has focused on diagonal approximations, low-rank representations, and stochastic variants of the quantum Fisher matrix to reduce this overhead while retaining some of the geometric benefits.

Secondly, we review the gradient-free methods, which mainly includes simultaneous perturbation stochastic approximation and evolutionary strategies.

\textbf{Simultaneous perturbation stochastic approximation (SPSA)} methods are stochastic algorithms particularly well-suited for high-dimensional parameter spaces and noisy optimization problems \cite{Spall1992}. Unlike gradient-based methods that require separate evaluations for each parameter, SPSA estimates the entire gradient vector using only two cost function evaluations per iteration, regardless of the number of parameters $p$. This remarkable property makes SPSA highly attractive for optimizing deep QNNs or large-scale variational quantum circuits where $p$ can reach thousands or more.

In SPSA, at each update step $t$, a random perturbation vector $\boldsymbol{\Delta}_t$ is generated, typically with each component drawn independently from a symmetric Bernoulli distribution (i.e., $\Delta_{t,i}\in\{+1,-1\}$ with equal probability). This perturbation vector is applied simultaneously to all parameters. The gradient is then approximated via the two-sided simultaneous perturbation estimate
\begin{equation}
\widetilde{\nabla}\mathcal{L}(\boldsymbol{\theta}_t)\approx\frac{\mathcal{L}(\boldsymbol{\theta}_t+c_t\boldsymbol{\Delta}_t)-\mathcal{L}(\boldsymbol{\theta}_t-c_t\boldsymbol{\Delta}_t)}{2c_t}\boldsymbol{\Delta}_t^{-1},    
\end{equation} where $c_t>0$ is a small shift magnitude that typically decays over iterations (e.g., $c_t=c/(t+1)^\gamma$ with $0<\gamma\leq 1$), and $\boldsymbol{\Delta}_t^{-1}$ denotes component-wise inversion. Due to the Bernoulli distribution, $\Delta_{t,i}^{-1}=\Delta_{t,i}$ because $+1$ and $-1$ are their own inverses. Intuitively, the random perturbation simultaneously explores all parameter directions, and the inner product structure of the estimate ensures that the expected value of $\widetilde{\nabla}\mathcal{L}(\boldsymbol{\theta}_t)$ approximates the true gradient $\nabla\mathcal{L}(\boldsymbol{\theta}_t)$ up to a bias that vanishes as $c_t \to 0$.

SPSA has been shown to be particularly well suited to QNNs and noisy intermediate scale quantum (NISQ) devices \cite{Kandala2017}. Its robustness to stochastic noise and its minimal circuit evaluation cost per iteration align well with the constraints of current quantum hardware, where circuit execution is expensive and measurement shots are limited. Moreover, SPSA requires no knowledge of the circuit's internal structure, gate derivatives, or analytic gradient rules, making it a truly blackbox optimizer that can be applied to any QNN, including those with nondifferentiable components.

However, the convergence and practical performance of SPSA depend critically on the choice of hyperparameters, particularly the learning rate schedule $\eta_t$ and the perturbation scale $c_t$. Poorly chosen schedules can lead to slow convergence, oscillations, or even divergence. More importantly, under dominant noise conditions, where the cost function evaluation is corrupted by significant hardware noise or finite-shot sampling error, SPSA often exhibits poor performance. The reason is that the gradient estimate involves a difference of two noisy cost values. When the signal-to-noise ratio is low, this difference is easily overwhelmed by statistical fluctuations. Furthermore, the random nature of the perturbation vector introduces additional variance into the gradient estimate, which can destabilize training in highly non-convex or flat landscapes. Consequently, while SPSA remains a valuable tool for initial exploration and for circuits where analytic gradients are unavailable, practitioners often turn to hybrid approaches or more sophisticated stochastic optimizers when high precision or robustness to noise is required.

\textbf{Evolutionary strategies}, such as \textbf{genetic algorithms (GAs)} and \textbf{particle swarm optimization (PSO)}, offer gradient-free optimization for QNNs via population-based search. Unlike gradient-dependent methods, these evolutionary techniques do not require cost function differentiability, making them particularly attractive for QNNs with non-smooth components, non-available analytic gradients, or measurement-based cost landscapes that exhibit discontinuities.

GAs evolve candidate QNN parameters or even entire circuit structures through stochastic operations inspired by biological evolution \cite{Lamata2018}. A typical GA maintains a population of candidate parameter vectors or circuit architectures. At each generation (one complete round of evaluation), individuals (single parameter vectors) are evaluated on the QNN cost function, and the fittest individuals are selected as parents. These parents then produce offspring via crossover (e.g., recombining parameter values from two successful candidates) and mutation (e.g., application of small random perturbations to parameters). The algorithm iterates through selection, crossover, and mutation cycles, gradually steering the population toward low cost regions of the parameter space. GAs are particularly well-suited for discrete or mixed optimization problems, such as discovering optimal gate sequences or circuit topologies, because crossover and mutation can operate directly on structural representations rather than continuous parameter values.

PSO, by contrast, updates each candidate (or ``particle'') based on a combination of its own best-known position and the global best position discovered by the entire swarm \cite{Li2024}. Each particle maintains a position vector $\boldsymbol{\theta}_i$ (representing QNN parameters) and a velocity vector $\boldsymbol{v}_i$. The update rules are
\begin{equation}
\begin{split}
\boldsymbol{v}_i&\leftarrow\omega\boldsymbol{v}_i+c_1\boldsymbol{r}_1\odot(\boldsymbol{\theta}_i^{\text{pbest}}-\boldsymbol{\theta}_i)+c_2 \boldsymbol{r}_2\odot(\boldsymbol{\theta}^{\text{gbest}}-\boldsymbol{\theta}_i),\\    
\boldsymbol{\theta}_i&\leftarrow \boldsymbol{\theta}_i+\boldsymbol{v}_i,    
\end{split}
\end{equation}
where $\omega$ is an inertia weight controlling momentum, $c_1$ and $c_2$ are cognitive and social acceleration coefficients, $\boldsymbol{r}_1, \boldsymbol{r}_2$ are random vectors drawn uniformly from $[0,1]$, and $\odot$ denotes element-wise multiplication. $\boldsymbol{\theta}_i^{\text{pbest}}$ denotes the personal best position of particle $i$ in PSO, which in QNN training corresponds to the best set of quantum circuit parameters that this specific candidate QNN has achieved so far, yielding the lowest cost function value across its own search trajectory. $\boldsymbol{\theta}^{\text{gbest}}$ denotes the global best position of the entire swarm in PSO, which in QNN training corresponds to the single best quantum circuit parameter configuration ever discovered by any candidate QNN across all particles and generations, representing the globally optimal solution that minimizes the QNN cost function for the given quantum machine learning task. This formulation enables rapid convergence with minimal hyperparameter tuning, as the swarm collectively balances local exploration with global exploitation.

Both GAs and PSO share several advantages for QNN optimization. First, their computational cost scales independently of circuit differentiability. They require only cost function evaluations, not gradients. This makes them applicable to any QNN, including those with non-differentiable quantum operations or hardware constraints. Second, both methods remain viable under noise. The population-based averaging across candidates naturally mitigates stochastic fluctuations in cost evaluations, and the inherent exploration in mutation (in GAs) or velocity perturbations (in PSO) helps escape local minima induced by noise. Third, these algorithms are inherently parallelizable, as each candidate's cost evaluation can be executed independently on separate quantum processors or simulators, offering nearlinear speedups with available resources.

However, evolutionary strategies also face limitations. The number of cost evaluations per generation scales with population size, which is usually of the order 10 to $10^3$. This can be prohibitively expensive for large QNNs or costly quantum hardware access. Moreover, convergence rates are often slower than gradient-based methods in smooth, low-noise regimes, as evolutionary algorithms rely on random variation rather than directed derivative information. Consequently, GAs and PSO are most appropriate for training the QNN where gradients are unavailable, unreliable, or expensive to compute, or where discrete structural optimization, such as finding optimal circuit architectures, is required alongside continuous parameter tuning.

Beyond the methods outlined above, several additional optimization techniques have been applied to QNN training, including the Nelder-Mead algorithm \cite{Nelder1965}, the L-BFGS-B algorithm \cite{Aminpour2024}, and others \cite{Bonet-Monroig2023}. The Nelder-Mead method is a derivative-free direct search algorithm that maintains a simplex of $p+1$ vertices, making it suitable for low-dimensional QNNs ($p\lesssim 10$) where gradient information is unavailable or unreliable. However, its convergence degrades significantly in high-dimensional parameter spaces. The L-BFGS-B algorithm, by contrast, is a quasi-Newton method that approximates the inverse Hessian using limited gradient history, achieving superlinear convergence when accurate gradients are available. Its bound constraints also allow restricting parameters to physically meaningful ranges, such as angular parameters within $[0,2\pi)$. However, both methods face challenges on noisy quantum hardware. Nelder-Mead can stagnate due to cost fluctuations, while L-BFGS-B suffers from inaccurate gradient estimates under finite-shot noise. Other derivative-free methods, such as COBYLA and trust-region approaches \cite{Bonet-Monroig2023}, have also been explored but share similar trade-offs. We do not go into further details here, as their core principles follow from classical optimization theory. The main difference is that QNN objective evaluations are far more expensive and noisy than their classical counterparts. 

\section{Typical Structures of QNNs}\label{Sec:Structure}
In this section, we summarize the {\it basic} structural requirements for QNNs. As the foundation, we start with a QNN that imposes no extra requirements beyond unitarity. We refer to this baseline architecture as a fully connected QNN (FCQNN), drawing a conceptual parallel to the fully connected classical NNs, which likewise represents the most basic and widely used structure in classical NNs. Then, building upon this foundation with additional structural requirements, we review several types of QNNs architectures, including QCNNs (together with QDNNs), EQNNs, QHMs/QBMs, and QRC.
\subsection{FCQNNs for Supervised Learning}\label{subsec:QCQNN}
As a baseline architecture, this type of QNN employs qubits as the ``neurons'' and unitary gates as the ``connections between neurons''. In principle, all qubits in such a network can be interconnected via these gates, analogous to how every neuron in a classical fully connected layer connects to all neurons in the subsequent layer. Hence, this architecture is therefore termed ``fully connected'' . Early discussions of FCQNNs were presented in Ref. \cite{Matsui2000}, and a typical FCQNN is sketched in Fig. \ref{FIG:2}(b). From Fig. \ref{FIG:2}(b), one can observe that an FCQNN establishes a parametric input-output relation, mapping an input quantum state to an output quantum state through a sequence of parameterized unitary transformations. Consequently, the model is broadly applicable to supervised learning tasks, where all data samples are accompanied by clearly defined labels.

\subsubsection{Learning Procedure}

The learning procedure using an FCQNN follows a typical pipeline, described as follows.

\textbf{Encoding.} First, all data samples of a given dataset must be encoded into quantum states represented by qubits. Numerous encoding strategies have been proposed in the literature \cite{Rebentrost2014,Lloyd2014,Giovannetti2008}. They includes amplitude encoding \cite{Araujo2021}, which maps classical data vectors into probability amplitudes of a quantum state, dense qubit encoding \cite{Larose2020}, which exploits entanglement to pack multiple classical features into a single qubit, and angle encoding, which embeds features as rotation angles of single-qubit gates. The choice of encoding scheme significantly impacts the expressivity and resource efficiency of the subsequent QNN processing.

\textbf{Processing.} Second, the encoded state $|x\rangle$ of a sample $x$ is processed by multiple layers of unitary gates, denoted by $U_i(\boldsymbol{\theta}_i)$ for $i = 1, \ldots, L$, where $\boldsymbol{\theta}_i$ is a parameter vector of the $i$-th layer. The output qubit state $|\psi_{\text{out}}\rangle$ can then be expressed as
\begin{equation}
|\psi_{\text{out}}\rangle = \prod_{i=L}^{1} U_i(\boldsymbol{\theta}_i) |x\rangle,    
\end{equation}
where the product is taken in descending order, meaning the rightmost gate is applied first. This layered structure mirrors the compositional depth of classical deep networks, enabling the model to learn hierarchical features from the input data.

\textbf{Measurement and loss evaluation.} Third, the deviation of the model output from the true label is assessed by measuring $|\psi_{\text{out}}\rangle$ with respect to a chosen basis or observable. The measurement outcome, denoted by $M(|\psi_{\text{out}}\rangle)$, is typically the expectation value of a Hermitian operator, e.g., a Pauli observable $Z$ on a designated qubit. The parameters of the FCQNN must then be optimized so that the measurement outcomes $\{M(|\psi_{\text{out}}\rangle)\}$ across all training samples become as close as possible to the given labels. The closeness is quantified by a loss function. For supervised learning, commonly used loss functions include the mean squared error, cross-entropy, and the informational metric employed in QNGD \cite{Wierichs2022} (see Section \ref{subsec:TM}). We do not delve into their further details here, as they follow principles analogous to classical supervised learning, albeit with the added complexity of quantum measurement statistics.

\subsubsection{Key Characteristics in Comparison to Classical NNs}

We now summarize the important characteristics of FCQNNs in reference to classical fully connected NNs, focusing on three aspects: data flow, training dynamics, and empirical performance.

\textbf{Data flow.} A fundamental characteristic of FCQNNs is that they process data encoded as quantum states and model patterns through layer-wise unitary transformations. Due to the properties of quantum mechanics 
\cite{Yang2020}, the encoded data samples are naturally normalized after the encoding procedure. Specifically, quantum states are always represented by unit vectors in Hilbert space, i.e., $\langle\psi|\psi\rangle=1$. This intrinsic normalization is quite different from the data flow in classical fully connected NNs, where activations can grow or shrink arbitrarily across layers and typically require explicit normalization layers (e.g., batch normalization) to stabilize training. In fact, normalization is widely adopted in modern classical NNs precisely because it stabilizes activations, mitigates vanishing or exploding gradients, and thereby accelerates convergence. The built-in normalization of quantum states thus offers an inherent advantage. Furthermore, qubits exhibit correlation phenomena during FCQNN execution that have no direct classical analogs, such as quantum entanglement \cite{Ballarin2023}, scrambled quantum information \cite{Shen2020}, and quantum phase transitions \cite{Zhang2024}. These effects potentially offer shortcuts for generating demanded output patterns, as entangled states can encode correlations that would require exponentially many classical bits to represent in general.

\textbf{Training dynamics.} The distinctive data flow of FCQNNs fundamentally affects their training dynamics, which also diverge appreciably from those of classical fully connected NNs. The specific training methods applicable to FCQNNs are among those introduced in Section \ref{subsec:TM}, including gradient-based techniques and gradient-free techniques. One might therefore expect that the training of FCQNNs could generally converge faster than classical counterparts, which is a fundamental demand in the research field of machine learning. However, empirical results are often not satisfactory, because the gradient of the loss function tends to vanish exponentially with the number of qubits and circuit depth, i.e., the barren plateau problem \cite{McClean2018}. In recent years, this direction has attracted substantial attention \cite{Sannia2024,Larocca2024,Friedrich2024}, as barren plateaus pose a major obstacle to scaling up QNNs for practical applications. A huge effort has been spent on three main trial directions: (i) defining better loss functions that exhibit favorable landscape geometry, (ii) changing the structures of QNNs, e.g., employing carefully designed ansatze or pre-training routines, and (iii) proposing new optimization methods that can navigate flat landscapes. Because this issue involves the specific structures of QNNs, we summarize the relevant progress of the research field in the end .

\textbf{Performance.} In specific tasks, the advantages of FCQNNs over classical fully connected NNs have been demonstrated. For example, the quantum nature of FCQNNs inherently admits an enhanced feature space, benefiting classification tasks in supervised learning \cite{Havlíček2019,Li2022}. This enhancement arises because quantum states live in a Hilbert space of dimension $2^n$ for $n$ qubits, which is exponentially larger than classical feature spaces of comparable bit length. Furthermore, analysis based on the Fisher information spectrum shows that a class of FCQNNs can achieve a considerably larger effective dimension than comparable feedforward networks \cite{Abbas2021}, indicating a superior representation capability. It means that these models can capture more complex patterns with fewer trainable parameters. By increasing the depth of the networks, FCQNNs can also be applied to deep learning tasks. For instance, numerical experiments have demonstrated that deep FCQNNs can achieve a higher accuracy rate in image classification compared to classical benchmarks \cite{Zhao2021b}. In general, FCQNNs fundamentally manipulate data within the quantum state space, so the learning procedure can be enhanced by advances in quantum computing, such as quantum parallelism \cite{Wu2025}, which allows simultaneous processing of multiple inputs or multiple components of a single input in superposition. This parallelism, when combined with clever encoding schemes, offers the potential for exponential speedups in certain learning tasks, although realizing such speedups on near-term hardware remains an open challenge.

\subsection{QCNNs for Pattern Recognitions}
One of the underlying reasons for the unsatisfactory performance of FCQNNs lies in their lack of nonlinear transformations on data. Meanwhile, the unitary gates in an ordinary FCQNN do not change the size of data encoded by qubits, making them inefficient for extracting the intrinsic patterns of data samples. A significant advance in addressing both shortcomings is the QCNNs, which introduces operations controlled by measurements in its structures.

\subsubsection{The Structure of a QCNN}
The QCNN structure \cite{Cong2019} is sketched in Fig. \ref{FIG:2}(c). It can be observed that unitary gates and measurement-based rotations in a QCNN are alternated layer-by-layer. The unitary layers consist of parameterized two-qubit gates, e.g., the combinations of CNOT gates and single-qubit rotations, that entangle neighboring qubits, analogous to the convolution filters in classical convolutional NNs. The measurement-rotation layers perform partial measurements on a subset of qubits, followed by conditional rotations, which effectively reduce the number of qubits while preserving relevant quantum information.

More formally, for an input quantum state $|\psi_{\text{in}}\rangle$ of an $n$-qubit Hilbert space $\mathcal{H}_n$. A QCNN architecture alternates between two types of layers,

\begin{itemize}
    \item \textbf{Convolutional layer (unitary operators):} A layer of parameterized two-qubit unitary gates $U(\boldsymbol{\theta})$ applied to neighboring qubit pairs. For a set of parameters $\boldsymbol{\theta}=\{\theta_1, \dots,\theta_m\}$, the unitary transformation takes the form
    \begin{equation}
    \mathcal{U}(\boldsymbol{\theta})=\bigotimes_{\text{pairs }(i,j)} U_{i,j}(\theta_{i,j}),    
    \end{equation}
    where $U_{i,j}(\theta_{i,j})$ acts on qubits $i$ and $j$, typically consisting of entangling gates (e.g., CNOT) and parameterized single-qubit rotations (e.g., $R_y(\theta)$ or $R_z(\theta)$). These layers process the quantum state without changing the number of qubits, analogous to feature extraction in classical convolutional NNs.
    
    \item \textbf{Pooling layer (measurement rotations):} A layer that reduces the qubit count via measurement-based operations. For a given set of qubits to be pooled, one performs a measurement in a specified basis (e.g., the Pauli $Z$ basis), yielding a classical outcome $m$ with probability $p_m=\text{Tr}[\Pi_m\rho\Pi_m^\dagger]$. Conditioned on $m$, a unitary rotation $R(m)$ is applied to the remaining qubits. Tracing out the measured qubits, the effective transformation on the reduced density matrix $\rho_{\text{r}}$ of the remaining qubits is given by
    \begin{equation}
    \rho_{\text{r}}=\mathcal{P}(\rho)=\sum_{m}R(m)\text{Tr}_{m}\left[\Pi_m \rho\Pi_m^\dagger\right]R(m)^\dagger.    
    \end{equation}
    This operation reduces the Hilbert space dimension from $2^n$ to $2^{n/p}$ (with integer $p\ge2$), analogous to pooling or downsampling in classical convolutional NNs.
\end{itemize}

The alternating structure of a QCNN can thus be represented as
\begin{equation}
\rho_{\text{out}}=\mathcal{U}_K(\boldsymbol{\theta}_K)\mathcal{P}_{K-1}\left(\cdots\mathcal{P}_2\left(\mathcal{U}_2(\boldsymbol{\theta}_2)\mathcal{P}_1\left(\mathcal{U}_1(\boldsymbol{\theta}_1) |\psi_{\text{in}}\rangle\langle\psi_{\text{in}}|\mathcal{U}^{\dagger}_1(\boldsymbol{\theta}_1)\right)\mathcal{U}^{\dagger}_2(\boldsymbol{\theta}_2)\right)\cdots\right)\mathcal{U}^{\dagger}_K(\boldsymbol{\theta}_K),    
\end{equation}
where $K$ is the number of alternating blocks, $\mathcal{U}_k$ denotes the convolutional unitary layer, and $\mathcal{P}_{K-1}$ denotes the pooling layer with measurement-based rotations. This structure is inspired by the multi-scale entanglement renormalization ansatz (MERA) \cite{Vidal2007,Evenbly2015,Grant2018}, a tensor network architecture originally developed for studying quantum many-body systems. In MERA, a similar alternating pattern of unitary disentanglers and isometric coarse-grainers efficiently captures scale-invariant quantum states. The QCNN adapts this idea for machine learning tasks by introducing trainable parameters and measurement-based pooling.

Notably, the layer terminology in QCNNs borrows from classical convolutional NNs but captures only the spirit of their functions. In classical convolutional NNs, convolutional layers can perform both feature transformation and data resizing simultaneously, although this is not a requirement. In QCNNs, however, convolutional layers only apply unitary transformations without altering data dimensions, while resizing is handled separately by measurement-based pooling layers.

\subsubsection{Advantages of QCNNs}\label{ssec:AdvQCNN}

The advantages of QCNNs can be summarized in two aspects.

\textbf{Scaling property and complexity.} As explained above, the structure of QCNNs inherits the benefit of the hierarchical MERA architecture, yielding a compact circuit whose depth scales only logarithmically with the input size. More precisely, for an $n$-qubit input, a QCNN typically has $O(\log n)$ depth and $O(n)$ trainable parameters \cite{Pesah2021}. This logarithmic scaling makes QCNNs exceptionally efficient for handling complex data structures, particularly binary classification tasks \cite{Oh2020}. More importantly, rigorous analysis of gradient scaling in QCNNs shows that the decay of gradient variance $\text{Var}[\partial_\theta\mathcal{L}]$ is at most polynomial in system size, implying the absence of barren plateaus \cite{Pesah2021,Mahmud2024}. This is a critical distinction from many other QNN architectures. Furthermore, the generalization error of QCNNs has been proven to scale as $T/N$, where $T$ is the number of trainable gates and $N$ is the size of the training dataset \cite{Caro2022}. This property suggests that QCNNs could be trained efficiently with smaller datasets compared to other types of QNNs, as the model complexity is well-controlled by the hierarchical structure.

\textbf{Performance and applications.} QCNNs have been shown to be robust against certain types of noise in current quantum devices \cite{Herrmann2022}, a vital property seldom found in other QNN architectures. This noise resilience is particularly important for the NISQ era, where hardware imperfections are unavoidable. For learning performance, QCNNs involving only two-qubit interactions have achieved excellent classification accuracy in tests on MNIST, Fashion-MNIST, and the Iris dataset \cite{Hur2022,MacCormack2022,Mahmud2024,Sun2024}, demonstrating faster convergence and higher classification accuracy over classical methods. In some cases, these advantages have been achieved with fewer trainable parameters than classical methods. More complicated deep learning tasks, such as multi-channel pattern learning \cite{Smaldone2023} and multiclass image classification \cite{Shi2024}, have also been tackled by QCNNs with different subroutines. Recent research has further revealed the potential of QCNNs for learning large-scale data with only a few qubits \cite{Barrué2025}.

It is worth mentioning that there are other architectures termed "QCNNs" in the development trajectories of QNNs \cite{Chen2017,Li2020,Liu2021,Wei2022}. Because these schemes are either conceptual or confined to specific quantum subroutines, we do not involve them in this review.

\subsubsection{A Similar Structure: QDNNs}
Apart from QCNNs, introducing measurements also leads to another QNN architecture called quantum ``dissipative'' neural networks, or QDNNs for short \cite{Beer2020,Zhao2021}, as sketched in Fig. \ref{FIG:2}(d). The term ``dissipative'' in QDNNs originates from the theory of open quantum systems. In that framework, a quantum system inevitably interacts with an external environment. The joint system evolves unitarily, but since the environment is not accessible, one traces over its degrees of freedom. This partial trace yields an effective, non-unitary evolution for the system's reduced density matrix, often described by a Lindblad master equation that captures dissipation, decoherence, and entropy increase. QDNNs borrow this principle. They deliberately introduce ancillary qubits as a controllable ``environment'', apply a parameterized unitary that entangles the data qubits with these ancillas, and then discard (trace out) the ancillas. The resulting transformation is a completely positive trace-preserving (CPTP) map, which can be viewed as parameterized version of physical evolution for an open quantum system.

Formally, a QDNN layer implements a CPTP map $\mathcal{E}$ on the input state $\rho_{\text{in}}$ by coupling it to $a$ ancillary qubits initialized in a reference state, typically $|0\rangle\langle 0|_{\text{anc}}^{\otimes a}$. A parameterized unitary $V(\boldsymbol{\phi})$ acts on the joint system, after which the ancillas are discarded via the partial trace, given by
\begin{equation}
\rho_{\text{out}}=\mathcal{E}(\rho_{\text{in}})=\operatorname{Tr}_{\text{anc}}\left[ V(\boldsymbol{\phi})\left(\rho_{\text{in}}\otimes |0\rangle\langle 0|_{\text{anc}}^{\otimes a}\right) V(\boldsymbol{\phi})^\dagger\right].    
\end{equation}
Here, $\rho_{\text{in}}$ is the density matrix of the input quantum state. $V(\boldsymbol{\phi})$ is a parameterized unitary operator acting on the joint Hilbert space of data qubits and ancillas. $\operatorname{Tr}_{\text{anc}}$ denotes the partial trace over the ancilla degrees of freedom.

This operation induces a non-unitary, dissipative transformation on the data qubits. The number of ancillas $a$ in each insertion serves as a hyper-parameter for tuning the effective ``information loss'' per layer, offering more flexibility than the fixed reduction scheme of QCNNs. By choosing different unitary ansatz for $V(\boldsymbol{\phi})$ and varying $a$, one can realize a wide family of CPTP maps, including strictly ``dissipative'' channels that drive the system toward a fixed point.

\textbf{Relations to QCNNs.} Both QDNNs and QCNNs introduce non-unitarity into QNNs to enable functionality beyond what purely unitary circuits can achieve, such as qubit reduction and enhanced representational power. At a functional level, both architectures reduce the effective Hilbert space dimension of the quantum state as information propagates through the network. This reduction is essential for extracting high-level features from the input data, analogously to downsampling or pooling operations in classical convolutional neural networks. Both approaches have a potential to be implemented on near-term quantum hardware with appropriate classical feedforward capabilities.

The fundamental distinction between QDNNs and QCNNs lies in how non-unitarity is introduced.

QCNNs achieve qubit reduction through projective measurements followed by conditional rotations. This process retains a clear interpretation rooted in MERA and actively uses measurement outcomes to determine subsequent operations. The transformation is not described as a CPTP map in the general sense. Instead, it is a measurement feedforward protocol.
    
QDNNs achieve non-unitary transformations through the open-system paradigm, which is inducing unitary coupling to ancillary qubits and then tracing them out. This operation is equivalent to applying multi-qubit gates followed by a partial trace over the ancilla qubits, as illustrated in Fig.~\ref{FIG:2}(d). Unlike the measurement-induced collapse in QCNNs, the dissipation here arises from discarding quantum information of the ancillas, which can be done without an irreversible projective measurement. It makes the whole process compatible with coherent quantum error correction and other advanced techniques.

\textbf{Expressivity and trainability trade-offs} From a general perspective, QDNNs has a better expressivity than  QCNNs do. This is because QDNNs can represent arbitrary CPTP maps, including but not limited to dissipative channels. However, this expressivity comes at a significant cost. It has been shown that QDNNs exhibit barren plateaus, and their trainability is not always guaranteed \cite{Sharma2022}. The introduction of ancillas expands the parameter space and thereby causes the gradient vanishing problem, especially as the network depth increases. In contrast, the gradient variance decays at most polynomially for QCNNs, due to their logarithmic depth of the networks. Consequently, QDNNs are less reliable than QCNNs for deep architectures or large-scale problems without careful initialization, structured ansatz, or specialized training techniques.

Nevertheless, the dissipative nature of QDNNs offers unique advantages for specific tasks. Because they naturally model open-system dynamics, QDNNs are well-suited for learning from noisy or incomplete data, simulating biological or physical systems where dissipation is intrinsic, and implementing noise-resilient learning protocols. Engineered dissipation can actively extract entropy from the system, potentially stabilizing training dynamics and preventing optimization from being trapped in poor local minima. These features make QDNNs a valuable architecture for targeted applications, despite their trainability challenges.

\subsection{EQNNs for Learning Datasets with Special Symmetries}
\subsubsection{The Basics of EQNNs}
In NNs, equivariance is the property where the output changes predictably when the input undergoes specific transformations \cite{Gerken2023}. More formally, a model $f$ is equivariant under a group $G$ if for all $g \in G$ and all inputs $x$,
\begin{equation}
f(g\cdot x)=g\cdot f(x),    
\end{equation}
meaning that applying a symmetry transformation to the input leads to the same transformation applied to the output. This property is especially important when learning data with inherent symmetries, such as the one obtained from Hamiltonian systems invariant under rotations or translations, or the objects in images whose recognition should be independent of position or orientation. It has already been shown that equivariant classical NNs exploit these symmetries to enhance performance, reducing sample complexity and improving generalization \cite{Zhong2023}.

For a QNN with a parameterized unitary $V(\boldsymbol{\theta})$ and an input state $|\psi\rangle$ encoding the data, equivariance requires that there exists a unitary representation $U(g)$ acting on the Hilbert space such that
\begin{equation}
V(\boldsymbol{\theta})U(g)|\psi\rangle=U'(g)V(\boldsymbol{\theta})|\psi\rangle,\quad\forall g \in G, 
\end{equation}
where $U'(g)$ is a representation on the output space, which may be the identity for invariant case. The QNNs with such a property is termed EQNNs \cite{Skolik2023,West2023}. EQNNs aim to achieve similar improvements on the foundation of existing QNN architectures such as FCQNNs, QCNNs, and others. In the previously introduced architectures, the unitary gates in FCQNNs and QCNNs are arbitrary, imposing no constraints beyond unitarity. EQNNs, by contrast, employ the same circuit topology but require the gates to be equivariant under specific symmetry transformations. When the symmetry transformation of the input and output state equals to each other, i.e., $U(g)=U'(g)$, a sufficient condition for achieving equivariance is 
\begin{equation}
[V(\boldsymbol{\theta}),U(g)]=0,\quad\forall g\in G,   
\end{equation}
i.e., the parameterized operations commutes with the symmetry representation. If, in addition, the measurement observable $O$ satisfies $U(g)OU(g)^\dagger=O$, then the entire QNN output becomes $G$-invariant or $G$-equivariant depending on the choice of output representation \cite{Nguyen2024}. In some contexts, the approach is also known as geometric quantum machine learning \cite{Tam2026,Das2024b,Schatzki2024}, emphasizing on the learning process aided by the underlying geometry of the state space. The key insight is that designing architectures that encode the symmetries of the problem at hand can mitigate fundamental challenges such as barren plateaus, poor local minima, and high sample complexity \cite{Bultrini2023,Tüysüz2024}. By constraining the quantum circuit to commute with the symmetry group representation, the effective Hilbert space is restricted to a symmetric subspace, which reduces the dimensionality of the optimization landscape while preserving expressive power for symmetric tasks. The specific advantages of introducing these equivariance constraints are summarized below.

\subsubsection{Training Properties}

As explained throughout this review, QNNs generally suffer from excessive local minima and barren plateaus. Exponentially flat cost function landscapes that render gradient-based optimization ineffective for large-scale problems. EQNNs have emerged as promising candidates that can mitigate these issues \cite{Nguyen2024}. For instance, an analytical study of permutation-EQNNs (equivariant under the symmetric group $S_n$) proves that they do not suffer from barren plateaus, quickly reach overparametrization, and generalize well from small amounts of data \cite{Schatzki2024}. To see why, consider the gradient variance $\text{Var}[\partial_\theta\mathcal{L}]$ for a generic QNN, which typically decays exponentially with the number of qubits $n$, i.e., $\text{Var}[\partial_\theta\mathcal{L}] \sim O\left(1/2^n\right)$. For an EQNN that respects the symmetry group $G$, the effective Hilbert space is restricted to the symmetric subspace, which is usually exponentially smaller than the full space. Consequently, the gradient variance scales as
the inverse of the dimension of the subspace, which is usually a polynomial in $n$ for many symmetry groups, thereby eliminating barren plateaus \cite{Schatzki2024}. These theoretical guarantees represent a significant advancement, as they provide the first rigorous evidence that certain QNN architectures can be provably trainable even at scale. 

Furthermore, recent research on resource-efficient equivariant QCNNs has shown that incorporating symmetry can also improve measurement efficiency. The split-parallelizing QCNN \cite{Chinzei2024,Chinzei2026}, for example, encodes general symmetries into the quantum circuit by splitting the computation at pooling layers, achieving an $O(n)$ improvement in measurement efficiency compared to conventional equivariant QCNNs while preserving trainability and the absence of barren plateaus. These advances suggest that equivariance not only aids trainability but also addresses the practical challenge of measurement cost in near-term quantum devices.

\subsubsection{Applications and Empirical Performance}

It has been reported that EQNNs achieve improved performance across a variety of tasks. For example, permutation-EQNNs have been proven applicable to learning heuristic rules for combinatorial optimization problems, outperforming QNNs without permutation equivariance on the same tasks \cite{Skolik2023}. Reflection-EQNNs have been shown to outperform generic ansatze on high-resolution image classification tasks \cite{West2023}. In a binary classification task, $Z_2 \times Z_2$-EQNNs outperform their classical counterparts, particularly in scenarios with fewer parameters and smaller training datasets \cite{Dong2024}, demonstrating that the inductive bias provided by equivariance can be especially valuable in data-scarce regimes. A key measure of this advantage is the reduction in sample complexity. For an equivariant model, the number of training samples required to achieve a given generalization error $\epsilon$ scales as
\begin{equation}
N_{\text{sample}}\sim\frac{\dim(\mathcal{H}_{\text{sym}})}{\epsilon^2},    
\end{equation}
where $\dim(\mathcal{H}_{\text{sym}})$ is the dimension of the symmetric Hilbert subspace. It is advantageous compared to the scaling $N_{\text{sample}} \sim 2^n/\epsilon^2$ for a non-equivariant QNN operating on the full Hilbert space \cite{Schatzki2024}. 

For graph classification tasks, such as those on the MNIST and Fashion-MNIST datasets interpreted as graphs, the permutation-EQNN with the layer structure of QCNNs, i.e., the permutation-equivariant QCNN, shows better average performance compared to EQNNs without convolutional layers \cite{Das2025}. This suggests that combining equivariance with hierarchical feature extraction yields a particularly powerful architecture for structured data. Additionally, recent work on rotational equivariant QNNs for point cloud processing demonstrates that cylindrical voxelization encoding can achieve comparable accuracy with smaller parameter scales than hardware-efficient ansatze, while also alleviating the barren plateau problem \cite{Li2024b,Park2025}. These findings reveal the adaptive relationship between encoding characteristics and model architecture, providing a potentially feasible solution for processing complex point clouds with large-scale qubits.

\subsubsection{Robustness under Hardware Noise}

As a necessary step toward practical deployment, the robustness of EQNNs under hardware noise has also been systematically analyzed \cite{Tüysüz2024}. Studies investigating the behavior of EQNN models in the presence of noise reveal that certain EQNNs can preserve equivariance under Pauli channels, such as the bit-flip or phase-flip errors. This preservation is not possible under the amplitude damping channel, which models energy relaxation. The degradation of equivariance under noise can be quantified by the symmetry-breaking error $\Delta(g)$ defined as
\begin{equation}
\Delta(g)=\left\|\mathcal{N}\circ\mathcal{E}\circ\mathcal{N}^{-1}(U(g)\rho U(g)^\dagger)-U(g) \mathcal{E}(\rho)U(g)^\dagger\right\|_1,    
\end{equation}
where $\mathcal{E}$ is the ideal EQNN channel, $\mathcal{N}$ is the noise channel, and $\|\cdot\|_1$ denotes the trace distance. For Pauli channels, $\Delta(g)=0$, meaning equivariance is preserved. For amplitude damping, $\Delta(g)$ grows linearly with the number of layers and noise strength, implying that deeper EQNNs are more vulnerable to decoherence \cite{Tüysüz2024}. Importantly, this line of research also provides mitigation strategies to enhance symmetry protection in EQNN models, validated through numerical simulations and hardware experiments up to 64 qubits. These findings are crucial for implementing EQNNs on noisy intermediate-scale quantum devices, where error rates are non-negligible.

\subsubsection{Open Challenges and Future Directions}

Because equivariance is a relatively young research topic for QNNs, there remain many open questions regarding the benchmarking and theoretical understanding of EQNNs. To date, only a limited set of examples, the majority of which are already tackled by equivariant classical NNs, have been handled by EQNNs. Problems such as how EQNNs perform on a broader range of tasks, how to design EQNNs for arbitrary symmetry groups beyond permutations and rotations, and how to understand their performance theoretically remain worth investigating. Additionally, recent work has questioned the conventional understanding of generalization in quantum machine learning, showing that state-of-the-art QNNs can accurately fit random states and random labeling of training data \cite{Schatzki2024}. This finding challenges existing notions of generalization based on complexity measures such as VC dimension \cite{GilFuster2024} or Rademacher complexity \cite{Bu2022}, and it remains an open question how equivariance affects this phenomenon. As the field matures, the development of more sophisticated theoretical tools and empirical benchmarks will be essential for realizing the full potential of EQNNs in practical applications.

\subsection{QHNs/QBMs for Unsupervised Learning}

QHNs \cite{Ma1993,Ma1995} and QBMs~\cite{Wiebe2014} are two QNN families built from Ising-type gates. Both exploit the equilibrium statistics of quantum Ising Hamiltonians, yet they differ in purpose and design. QHNs, introduced first, originally arise as a condensed matter model \cite{Ma1993,Xi1999,Xi2000} and operate as associative memories \cite{Nishimori1996,Meinhardt2020} from an informational perspective. QBMs, developed later, restrict the state distribution to a Boltzmann form in quantum regime \cite{Wiebe2014,Amin2018,Coopmans2024} and use controlled dissipation to learn complex and high-dimensional probability distributions. This generative capability renders QBMs effective for sampling tasks and unsupervised discovery of latent structure in unlabeled data, or unsupervised learning tasks. In classical machine learning theory, unsupervised learning refers to a paradigm in which a model is trained on input data without labeled outputs, aiming to uncover hidden patterns, features, or distributions intrinsic to the data itself, such as clustering, dimensionality reduction, or probabilistic modeling, without any external supervision signal.

Usually, QHNs and QBMs are graphically represented by qubits or spins with specific connections.  Mapping quantum evolutions to networks of unitary gates, the circuit of a QBM is shown in Fig. \ref{FIG:2}(f). The visible qubits in \ref{FIG:2}(f) are for data injection and readout, and the hidden qubits are for enriching the dynamical patterns without external manifestation. The circuit of a QHN is the same as Fig. \ref{FIG:2}(f), except that the two kinds of qubits are not distinguished in general. Both models are trained by measuring the statistical distance between the output and desired distributions. Two common measures are the quantum relative entropy,
\begin{equation}
D(\rho\|\rho')=\mathrm{Tr}\bigl[\rho(\log\rho-\log{\rho'})\bigr],
\end{equation}
where $\rho$ and $\rho'$ are the density operators of two quantum states, and the classical Kullback–Leibler divergence applied to measurement outcome probabilities,
\begin{equation}
D_{KL}(p_{\text{out}}\| p_{\text{target}})=\sum_{\mathbf{m}}p_{\text{target}}(\mathbf{m}) \log\frac{p_{\text{target}}(\mathbf{m})}{p_{\text{out}}(\mathbf{m})},
\end{equation}
where $\mathbf{m}$ denotes a bitstring obtained from computational basis measurements. $p_{\text{out}}(\mathbf{m})$ ($p_{\text{target}}(\mathbf{m})$) is the probability of measuring $\mathbf{m}$ in the outputs (the given dataset).

\subsubsection{Researches on QHNs}
The researches on QHNs can be summarized into two directions.

\textbf{Statistical characters.} Using Trotter decompositions and replica methods, the works \cite{Ma1993,Xi1999,Nishimori1996} on equilibrium locate the critical storage capacity,
\begin{equation}
\alpha_c=P_c/N,
\end{equation}
where $P_c$ is the maximum number of patterns that can be stored reliably, $N$ is the number of qubits, and $\alpha_c$ is found to be comparable to or larger than the classical Hopfield limit $\alpha_c^{\text{(classical)}} \approx 0.138$ \cite{Amit1985}, depending on the transverse field strength and disorder. This capacity is further refined by the extensions \cite{Xi2000,Bödeker2023b} to random fields, which show that disorder can enhance or degrade memory performance depending on the noise regime. The open-system treatments \cite{Fiorelli2022,Rotondo2018} reveal how noise distorts the distribution landscape, confirmed by recent simulations \cite{Torres2024}. Specifically, the master equation for the density matrix,
\begin{equation}
\frac{d\rho}{dt}=-i[H_{\text{Ising}},\rho]+\sum_k\gamma_k\left(L_k\rho L_k^\dagger-\frac{1}{2}\{L_k^{\dagger}L_k,\rho\}\right),
\end{equation}
where $L_k$s are jump operators modeling decoherence, shows that the retrieval error increases exponentially with the dissipation rate $\gamma_k$ beyond a threshold. The studies of generalized $k$-body \cite{Seki2015} and XY interactions \cite{Shcherbina2020} broaden the phase diagram of QHNs, collectively showing essential conditions for robust associative memory in noisy quantum regimes.

\textbf{Learning \& algorithmic properties.} The investigations for this aspect have converged on the topics of Hebbian rules and variational optimization. The works \cite{Liu2020b,Rebentrost2018,Miller2021} have demonstrated that the quantum Hebbian rule, which embeds patterns directly into the Ising-coupling matrix,
\begin{equation}
J_{ij} = \frac{1}{N} \sum_{\mu=1}^{P} \xi_i^\mu \xi_j^\mu,
\end{equation}
enables extensively fewer parameter counts than those required by classical counterparts. Unlike classical Hopfield networks that require $O(N^2)$ parameters, QHNs achieve similar or better storage with the same scaling but with a much lower effective constant due to quantum superposition. Extending this principle, Ref. \cite{Hariharasitaraman2024} couples the rule to cryptographic key-recovery tasks, showing that the learned couplings generalize across unseen cipher challenges by exploiting quantum tunneling between local minima. Refs. \cite{Srivastava2014,Srivastava2015} lift the Hebbian paradigm to hierarchical qudit architectures. It reveals that the pattern-association rates $R_{\text{assoc}}$ scale linearly with the qudit dimension $d$, expressed by $R_{\text{assoc}}\propto d\cdot\log(P)$, where $P$ is the number of stored patterns. This indicates higher-dimensional qudits offer exponential advantages in memory density compared to qubits. These works establish QHNs as data-efficient and high-fidelity associative memories. They open a direct route toward unsupervised learning of latent structure in noisy, unlabeled quantum datasets.

\subsubsection{Researches on QBMs}
The research on QBMs is also summarized into two directions.

\textbf{Algorithmic \& theoretical advances.} The theoretical advances of QBMs are mainly on sampling properties. It has been shown that QBMs enable polynomial-time Gibbs-state preparation without contrastive divergence \cite{Wiebe2014}. The quantum Boltzmann distribution is given by
\begin{equation}
\rho_{\mathrm{QBM}}=\frac{e^{-\beta{H}_{\mathrm{eff}}}}{Z},~Z=\mathrm{Tr}[e^{-\beta{H}_{\mathrm{eff}}}],
\end{equation}
with the effective Hamiltonian typically taking the form
\begin{equation}
{H}_{\mathrm{eff}}=\sum_i a_i{Z}_i+\sum_{i,j}W_{ij}{Z}_i{Z}_j+\sum_k b_k{X}_k,
\end{equation}
where $a_i$ are visible biases, $W_{ij}$ are couplings (often restricted to a bipartite structure), and $b_k$ are transverse fields on hidden units. The absence of a sign problem in this transverse-field Ising form allows efficient sampling. $\beta$ is the inverse temperature in the corresponding physical background, which can be viewed as a hyperparameter in QBMs. Also, the training of QBMs can be achieved by methods such as native quantum sampling \cite{Amin2018} and relative-entropy optimization \cite{Coopmans2024}, which evades barren plateaus at polynomial sample cost. The gradient of log-likelihood with respect to a parameter $\theta$ reads
\begin{equation}
\frac{\partial\mathcal{L}}{\partial\theta}=\sum_{\textbf{m}}p_{\text{target}}\left(\frac{\mathrm{Tr}[\Lambda_{\textbf{m}}\partial_{\theta}e^{-\beta{H}_{\mathrm{eff}}}]}{\mathrm{Tr}[\Lambda_{\textbf{m}}e^{-\beta{H}_{\mathrm{eff}}}]}-\frac{\mathrm{Tr}[\partial_{\theta}e^{-\beta{H}_{\mathrm{eff}}}]}{\mathrm{Tr}[e^{-\beta{H}_{\mathrm{eff}}}]}\right),
\end{equation}
where $\Lambda_{\textbf{m}}$ is the clamped operator given by the tensor product of $|\textbf{m}\rangle\langle\textbf{m}|$ and the identity within the space of the hidden nodes. It has been noted that since $H_{\text{eff}}$ and $\partial_{\theta}H$ are now matrices that do not commute, therefore expectations of $\partial_{\theta}H$ are not trivially obtained as in the classical case \cite{Amin2018}.

\textbf{Application.} QBMs have been applied to simulations, sampling, controls, and generative modeling. For example, they are used for investigating the finite-size scaling of phase transitions via Hilbert-space truncation \cite{Khalid2022}. In this context, the scaling of the correlation length near the critical point can be extracted from QBM-learned distributions. The negative-phase estimation of cybersecurity data based on restricted QBMs can be largely improved with the aid of a quantum annealer \cite{Moro2023}, achieving an order-of-magnitude reduction in sampling time compared to classical Markov chain Monte Carlo. Besides, an architecture involving QBMs is proven to learn low-rank quantum or classical states with fewer measurements \cite{Huijgen2024,Coopmans2023}. Specifically, the number of required measurement settings scales as $O(r\log N)$ for a state of rank $r$, sublinear in the Hilbert space dimension $N$. Extending to the issue of controls, a hybrid actor-critic reinforcement-learning agent with clamped QBMs has been deployed for beam-line optimization at CERN \cite{Schenk2024}, where the QBM acts as a generative model of beam configurations, accelerating policy search by two orders of magnitude. As subroutines, QBMs can be adopted as the generative cores for quantum-autoencoder schemes, which achieve competitive performance on MNIST via quantum Monte-Carlo training \cite{Khoshaman2018} or a fast Boltzmann sampler supported by D-Wave \cite{Winci2020}. In the latter, the sampling rate reaches up to thousands of effective samples per second on a 2048-qubit annealer, demonstrating practical viability for unsupervised learning tasks at scale.

\subsection{QRC for Temporal Pattern Learning}
\subsubsection{The Basics of QRC}
In classical machine learning, reservoir computing is a framework for recurrent NN where a fixed, randomly connected dynamical system is employed as the ``reservoir''. During the learning process, all the weights are fixed, and only a linear readout layer is trained \cite{Jaeger2001,Maass2002}. This approach avoids the computational cost and vanishing/exploding gradient problems of conventional recurrent NNs, which require full backpropagation through time to update all recurrent weights~\cite{Lukosevicius2009}. It leverages the properties of nonlinear dynamics and fading memory to process temporal data efficiently, making it particularly suitable for time-series prediction and sequence learning tasks~\cite{Grezes2025}. 

This principle directly inspires QRC, where a fixed quantum circuit or Hamiltonian serves as the reservoir, and only few-qubit measurements are learned. Similar to paradigm of classical reservoir computing, QRC maps input sequences into a high-dimensional Hilbert space where only a linear readout layer is learned \cite{Fujii2017}. QRC can also be given in circuit form, as shown in Fig. \ref{FIG:2}(f). From the figure, it can be seen that the data of a sequence is encoded by partial qubits which are measured and re-initialized after evolving under the same set of unitary gates. QRC is usually thought to has built-in memory generated by coherent evolution and dissipation, and can be trained by single-qubit or few-qubit measurements. Contrary to classical reservoir computing, the key spirit of QRC is introducing dynamical data of quantum systems to learn temporal data, which turns out to be effective in certain cases.

The general evolution of a QRC can be described by a discrete-time map. At each time step $t$, an input vector $\mathbf{s}_t\in\mathbb{R}^m$ is encoded into a set of input qubits. The reservoir state, represented by a density matrix $\rho_t$, evolves under a fixed $U_{R}$. After tracing out the input qubits, the updated reservoir state becomes
\begin{equation}
{\rho}_{t+1}=\mathcal{T}({\rho}_t,\mathbf{s}_t)=\mathrm{Tr}_{\text{in}}\left[{U}_{R}\left( {\rho}_t\otimes|\mathbf{s}_t\rangle\langle\mathbf{s}_t|\right){U}^\dagger_{R}\right],
\end{equation}
where $\mathcal{T}$ denotes the reservoir state transition map. In some cases, $U_{R}$ could be set to depend on $\mathbf{s}_t$. The readout layer then estimates a target output via a linear combination of measured observables,
\begin{equation}
{y}_t=\sum_{k=1}^{K}w_k\langle{O}_k\rangle_t+b,~\langle{O}_k\rangle_t=\mathrm{Tr}[{O}_k{\rho}_t],
\end{equation}
where $\{w_k\}_{k=1}^K$ and $b$ are the only trainable parameters. ${O}_k$ denotes the measurement operator on the $k$-th qubit. The training typically minimizes the mean squared error over a temporal sequence of length $T$,
\begin{equation}
\mathcal{L}=\frac{1}{T}\sum_{t=1}^{T}\left({y}_t-y_t^{\text{target}}\right)^2.
\end{equation}

The QRC framework offers several distinct advantages over classical reservoir computing. First, the Hilbert space dimension grows exponentially with the number of qubits $N$, providing a vast feature space even for moderate $N$. Second, quantum coherence and entanglement can serve as computational resources, the analogue of which has seldom been considered in the  classical cases. Third, the reservoir dynamics can be realized with fixed, untrained quantum circuits or natural Hamiltonian evolutions, eliminating the need for costly backpropagation through time. 

\subsubsection{The Investigations on QRC}
The investigations on QRC can be summarized into two categories. One for the properties of QRC itself, and the other for its applications.

\textbf{Properties of QRC.} Subsequent studies have dissected the unique properties of QRC, such as the unification of entanglement detection and nonlinear function approximation \cite{Ghosh2019}, the effect of thermalization transition on learning performance \cite{Martínez-Peña2021}, and the information processing capacity relying on quantum coherence \cite{Palacios2024}. A key quantity characterizing QRC performance is the memory capacity, which quantifies how much information from past inputs is preserved in the reservoir state. The linear memory capacity is defined as \cite{Fujii2017,Martínez-Peña2021}
\begin{equation}
C=\sum_{\tau=1}^{\infty}C_\tau,~
C_\tau=\frac{\mathrm{Cov}^2\bigl(y(t),s(t-\tau)\bigr)}{\sigma_y^2\sigma_s^2},
\end{equation}
where $C_\tau$ measures the contribution of the input from time $t-\tau$ to the output at time $t$, and the covariance and variances are computed over the temporal sequence. More generally, the information processing capacity (IPC) \cite{Martínez-Peña2020} extends this concept to nonlinear transformations of past inputs,
\begin{equation}
\mathrm{IPC}=\sum_{\tau=1}^{\infty}\sum_{d=1}^{\infty}C_{\tau,d},~ 
C_{\tau,d}=\frac{\mathrm{Cov}^2\bigl(y(t),s^d(t-\tau)\bigr)}{\sigma_y^2\sigma_{s^d}^2},
\end{equation}
where $d$ is dimension of input data. Recent work on many-body QRC based on the transverse-field Ising model has revealed a monotonic relationship between reservoir performance and quantum coherence, with coherence playing an essential role in the ergodic regime of the reservoir dynamics \cite{Palacios2024}. It has also been shown that quantum correlations and entanglement contribute to improved IPC, though these quantum effects are increasingly suppressed by decoherence and measurement noise \cite{Palacios2024}. In the Jaynes–Cummings model, QRC exhibits an unusual property. The nonlinear memory capacity of the QRC based on such a model can exceed linear memory capacity, which is a signature of genuine quantum nonlinearity that is absent in classical reservoirs \cite{Das2025b}.

Surprisingly, the role of dissipation has also been inverted from nuisance to necessity for QRC in cases requiring fading memory. The echo-state property ensures that the reservoir state asymptotically depends only on the recent input history, not on the initial condition, expressed by
\begin{equation}
\lim_{t\to\infty}\left\|{\rho}_t(\{\mathbf{s}\},{\rho}_0)-{\rho}_t(\{\mathbf{s}\}, {\rho}_0')\right\|_1=0,
\end{equation}
where $\|\cdot\|_1$ is the trace norm. This property is essential for robust temporal processing. For example, it has been shown that natural decoherence furnishes the echo-state and fading-memory properties \cite{Suzuki2022}, and a continuous-dissipation model introduced was proven to be universal for fading-memory maps \cite{Sannia2022}. Furthermore, modified versions of QRC have also been given by introducing nonlinear evolution \cite{Wang2024} or nonlinear vector autoregression \cite{Sornsaeng2024,Wang2025}, guaranteeing improvement in prediction tasks. Recent theoretical work has established that QRC can be universal under suitable conditions, with the minimal sufficient conditions for universality in both classical and quantum reservoir computing being precisely the fading memory property and a polynomial algebra structure of the reservoir's associated functionals, as proven via the Stone-Weierstrass theorem \cite{Monzani2024}. This unified framework further reveals that spatial multiplexing can serve as a computational resource for quantum systems \cite{Monzani2024}.

Another line of research focuses on employing QRCs for quantum information processing. Unlike classical temporal series prediction, the analysis on quantum states is usually more complicated in the classical computing frameworks, yet can be directly produced by QRCs. It has been demonstrated that QRCs can qualitatively recognize quantum entanglement and quantitatively estimate nonlinear functions of input quantum states, such as logarithmic negativity, von Neumann entropy, purity, and $\mathrm{Tr}(\rho^n)$ for arbitrary $n$, with high precision using only a single trained readout observable \cite{Ghosh2019}. This capability suggests that QRC can simplify generic quantum experiments where multiple observables would otherwise be required.

\textbf{Applications.} QRC is reported to achieve 99\% classification accuracy on a temporal waveform task using only nine measured basis states, matching results from classical reservoirs with 24 nodes \cite{Dudas2023}, as verified by coupled superconducting oscillators. It has been shown that QRC can replicate cognitive tasks such as multitasking and long-term memory, exploiting quantum many-body scars and further verified by arrays of Rydberg atoms \cite{Bravo2022}. In the Rydberg platform, the different interatomic species interactions naturally encode both inhibitory and excitatory connections, while many-body quantum scars provide a mechanism for long-term memory through weak ergodicity breaking \cite{Bravo2022}. Beyond classical benchmarks, QRC has been used for inherently quantum duties, such as extracting logarithmic negativity and purity of past input states \cite{Nokkala2024}. The quantum reservoir processing framework has been extended to estimate these quantities simultaneously, with a single reservoir producing a multi-dimensional output vector where each component estimates a distinct physical parameter \cite{Li2024c}.

In time-series prediction, QRC demonstrates superior forecasting of chaotic dynamics compared to linear autoregressive models. For the Mackey–Glass system (a canonical benchmark for chaotic time-series prediction), QRC based on both the full Jaynes–Cummings model and its dispersive limit achieves prediction accuracy comparable to classical reservoir computing while using substantially fewer physical resources \cite{Das2025b}. The normalized root mean square error (NRMSE) for short-term prediction can be reduced by up to an order of magnitude relative to naive linear models. Even more remarkably, minimal spin–boson configurations in the Jaynes–Cummings model, when combined with time multiplexing, exhibit strong nonlinear memory and expressive power, overcoming the limitations of equivalent qubit pairs \cite{Das2025b}. This opens pathways towards tunable, high-performance quantum machine learning architectures based on hybrid qubit-boson systems.

Technologically, QRC has been implemented across diverse physical platforms, including superconducting qubits \cite{Monzani2024b}, trapped ions, Rydberg atom arrays \cite{AraizaBravo2022}, and photonic systems \cite{Gyurik2025}. These implementations share a common operational principle: the natural Hamiltonian dynamics of a many-body quantum system serve as the reservoir, with inputs injected via local driving fields or state preparation, and outputs read out via few-qubit measurements \cite{Gyurik2025}. The fixed, non-trainable nature of the reservoir makes QRC particularly well-suited for NISQ devices, where deep variational circuits are hindered by decoherence and barren plateaus \cite{Monzani2024b}. Indeed, dissipation—traditionally viewed as a nuisance—can be harnessed as a computational resource, with non-unital noise (e.g., amplitude damping) providing the fading memory property essential for temporal processing while simultaneously enriching the reservoir dynamics \cite{Monzani2024b}. As the field matures, QRC is increasingly recognized not merely as a machine learning technique, but as a general paradigm for harnessing quantum dynamics for information processing tasks that require temporal memory and nonlinear transformation \cite{Gyurik2025}.

\section{Composite QNN Architectures for Advanced Learning Tasks}\label{Sec:CompQNN}
To address more complicated learning tasks, such as multi-scale graph inference, domain-adaptive time-series prediction, or cross-modal learning, basic QNN structures must be combined into deeper, composite architectures. This gives rise to three advanced paradigms: quantum reinforcement learning (QRL), quantum generative learning (QGL), and quantum transfer learning (QTL). Each paradigm faces a fundamental constraint that shapes the required QNN design, as elaborated below.

In reinforcement learning, direct observation of optimal actions cannot be mapped to a simple loss function. Instead, the agent must infer good behavior solely from reward signals received after interacting with the environment. This demands a QNN that supports \emph{interactive adaptation}, which specifically includes processing sequential observations, maintaining memory of past actions, and performing rapid updates based on sparse feedback. QNNs based on quantum dynamics, such as QHNs, QBMs, and QRC, naturally provide the required memory and temporal processing. Additionally, QDNNs offer noise resilience for learning on imperfect hardware, while EQNNs ensure consistent behavior under state transformations. These properties also enable the application of these two QNN families to QRL tasks.

In generative learning, the true data distribution is unknown. One can only sample from it, without direct access to its underlying probability density. The task is to learn a model that produces new samples indistinguishable from real ones. This demands a QNN that supports \emph{adversarial synthesis}, which typically involves the collective updating of two competing networks, called a generator and a discriminator. Based on these requirements, FCQNNs, which provide the expressivity needed for generating complex distributions, can serve as generators in QGL. Moreover, QCNNs, QDNNs, and EQNNs can also be applied to this learning paradigm, particularly when training under hardware noise or enforcement of physical symmetries is especially required.

In transfer learning, labeled data for the target task is scarce, while abundant data exists for a related source task. One cannot directly train a large model on an insufficient target domain. This demands a QNN that supports \emph{knowledge reuse}, which typically involves adapting parameters of the target NN with the help of partially pre-trained features. The criterion for assessing whether pre-trained features are useful is known as alignment. With this in mind, QNNs based on quantum dynamics are ideal for transferring temporal patterns across domains, while FCQNNs and QCNNs serve as flexible feature extractors. QDNNs enable robust transfer despite domain shifts, and EQNNs ensure that learned symmetries transfer meaningfully.

To align the core properties of QNNs with the learning needs described above, each basic QNN type embodies a core spirit that matches specific requirements, summarized as follows,

\begin{itemize}
\item \textbf{FCQNNs} — \emph{Enhanced expressivity}: suitable as baselines for any paradigm due to their lack of structural constraints.
\item \textbf{QCNNs and QDNNs} — \emph{Locality and noise resilience}: ideal for processing structured data and enabling robust learning under hardware imperfections.
\item \textbf{EQNNs} — \emph{Symmetry as inductive bias}: essential when data transformations preserve certain symmetries, improving sample efficiency and generalization.
\item \textbf{QNNs based on quantum dynamics (QHNs, QBMs, QRC)} — \emph{Memory through dynamics}: naturally suited for temporal and sequential tasks that require retaining information over time.
\end{itemize}

\begin{figure}\centering
\includegraphics[width=.82\textwidth]{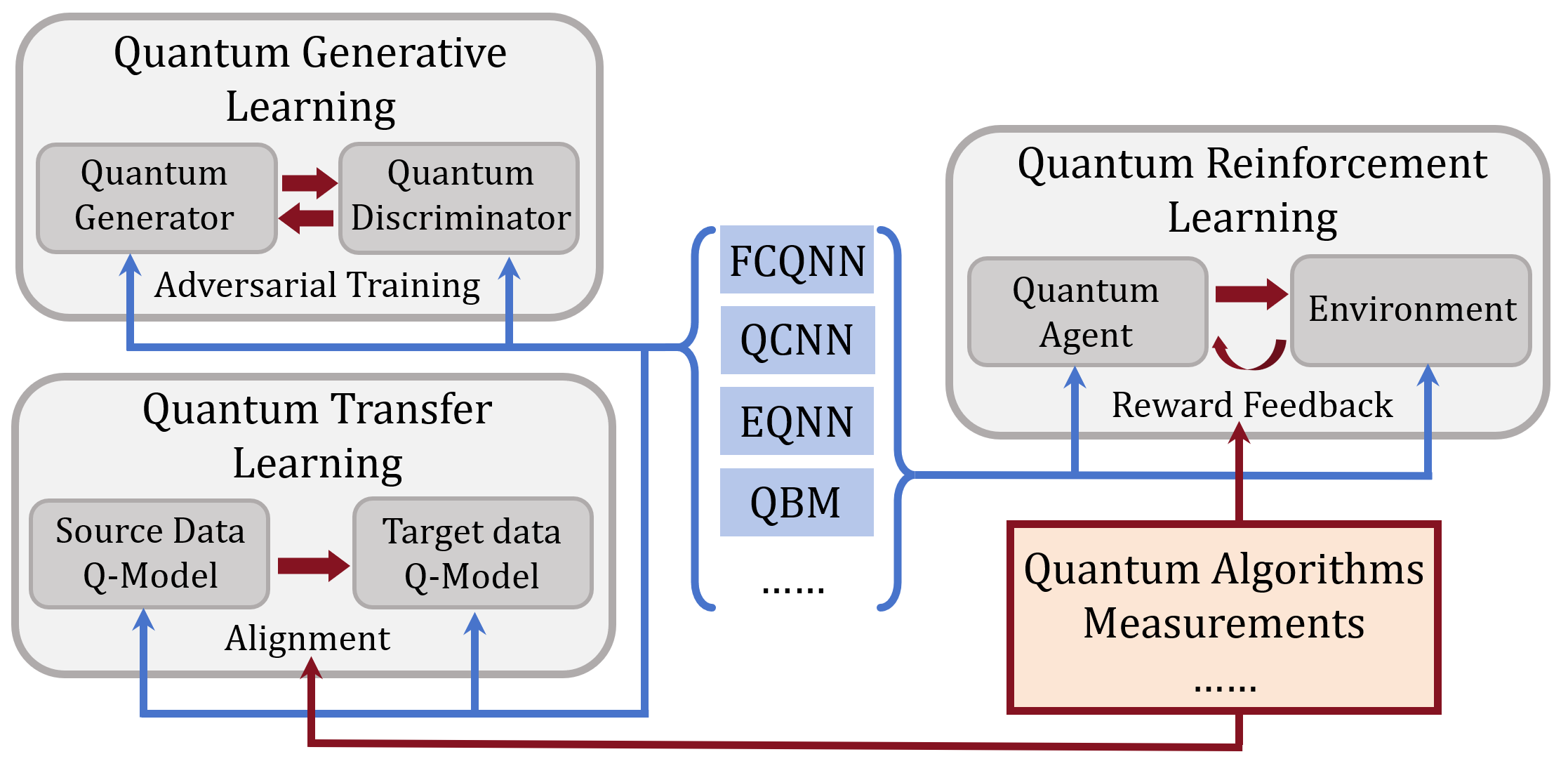}
\caption{A graphic illustration of the core requirements of QRL, QGL, and QTL. The roles of the basic types of the QNN for those learning paradigms are also illustrated, as indicated by blue arrows. The quantum operations on states, such as applying quantum algorithms or performing measurements, can be employed as a quantum version of alignment and reward feedback, as indicated by red arrows. combine into more complex architectures for these learning tasks.}\label{FIG:4}
\end{figure}

A graphic illustration of the general strategies is shown in Fig. \ref{FIG:4}. By aligning these architectural spirits with the respective constraints of each learning paradigm, composite QNNs can exhibit unique advantages in generalization, robustness, and sample efficiency. Detailed treatments of each paradigm appear in the following subsections.

\subsection{QNNs with Agents for QRL}

\begin{figure}\centering
\includegraphics[width=.69\textwidth]{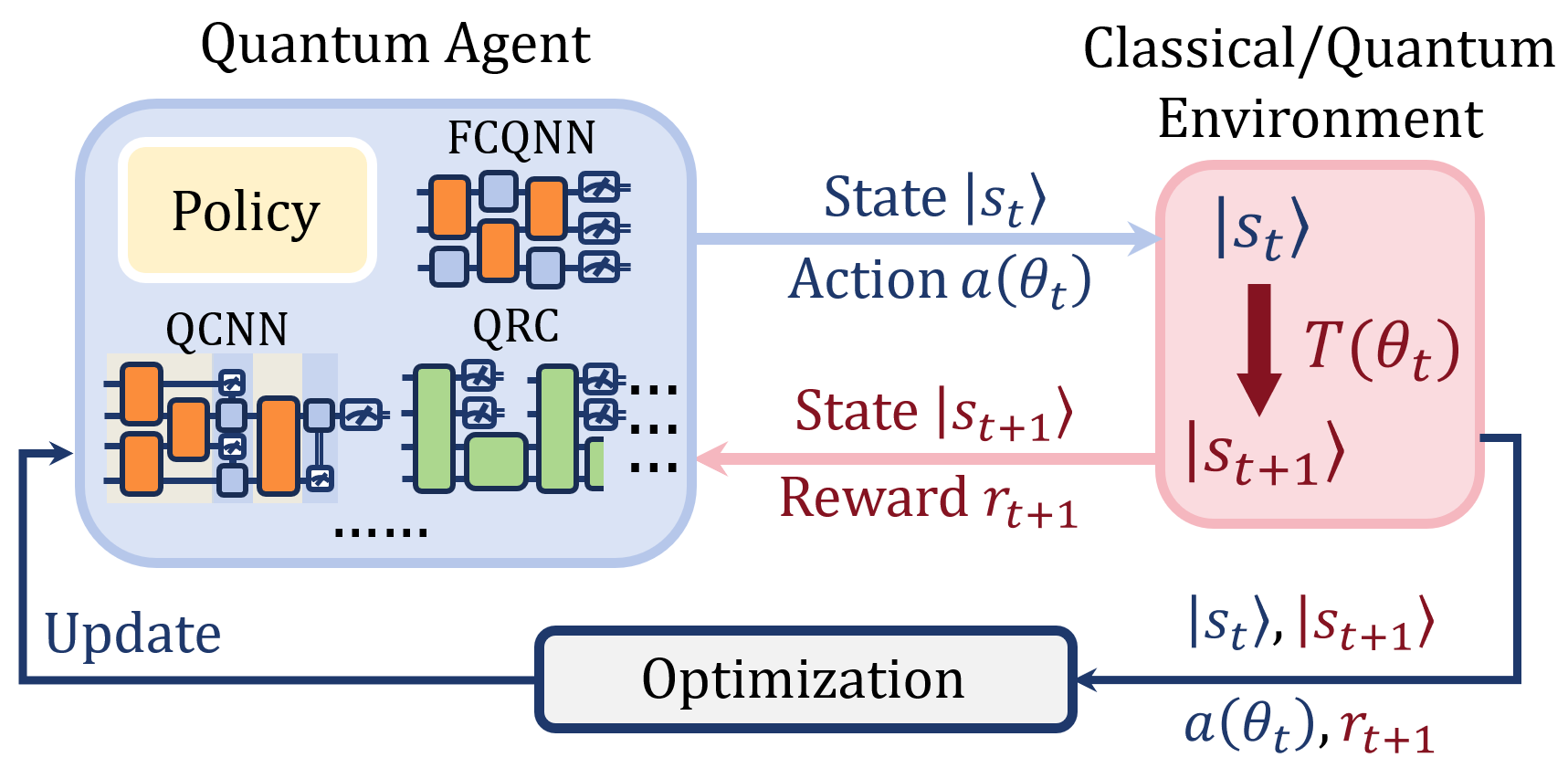}
\caption{Schematic of a QTL framework. A QNN-based agent interacts with an environment by observing a state $s_t$, selecting an action $a_t$ according to its policy, and receiving a reward $r_t$ as the environment transitions to a new state $s_{t+1}$ with probability $p(s_{t+1}|s_t,a_t)$. After collecting a set of transitions $(s_t,a_t,r_t,s_{t+1})$, the agent is updated using a suitable optimization algorithm.}\label{FIG:5}
\end{figure}

Combining QNNs under certain requirements enables the construction of composite networks for QRL, the quantum counterpart of classical reinforcement learning. Unlike supervised learning, where labels are precisely provided for all data points, and unsupervised learning, where labels are completely absent, reinforcement learning extracts useful information (often termed "rewards") through cumulative interaction with an environment. The basic scheme of QRL follows the same principle, as sketched in Fig.~\ref{FIG:5}. By employing a QNN, e.g., an FCQNN, QCNN, or others, as an agent, interaction with a classical or quantum environment enables this form of quantum machine learning. In the reinforcement learning literature, the agent is also commonly referred to as a \emph{policy}, emphasizing the mapping from environmental states to actions. Beyond QNN-based agents, multiple alternative pathways have been explored toward QRL \cite{Paparo2014,Rajagopal2021,Flamini2024,Wang2021b,Cherrat2023}. For a broader overview of these directions, see Ref. \cite{Meyer2022}.

\subsubsection{Core Methods}
The heart of QNN-based QRL lies in the structure of the agent and its interaction rules with the environment. Methods for constructing such agents with QNNs can be categorized into three frameworks: policy-based methods, value-based methods, and actor-critic frameworks.

\textbf{Policy-based methods.} In classical machine learning, such methods directly parameterize the policy. Extending the concept to QRL, QNNs can also employed as a model for the policy. A QNN outputs a probability distribution over actions given the current state. The objective is to maximize the expected cumulative reward, denoted as $J(\theta)=\mathbb{E}_{\tau\sim\pi_\theta}[R(\tau)]$, where $\pi_\theta$ represents the policy parameterized by $\theta$. $\mathbb{E}$ means taking the expectation according to certain conditions, and the same below. A policy here is a probability distribution over actions conditioned on the current state. The trajectory $\tau=(s_0,a_0,r_0,s_1,a_1,r_1,\dots)$ is generated by sampling actions according to the policy, i.e. $a_t \sim\pi_\theta(\cdot|s_t)$, while the environment responds with rewards and subsequent states according to its transition dynamics. To optimize $J(\theta)$, one computes the policy gradient, which for a stochastic policy takes the form
\begin{equation}
\nabla_\theta J(\theta)=\mathbb{E}_{\tau\sim\pi_\theta}\left[\sum_{t=0}^{T} \nabla_\theta\log\pi_\theta(a_t|s_t)R(\tau)\right].
\end{equation}
This estimator, known as REINFORCE, was introduced by Williams in 1992 and forms the foundation for many policy-gradient methods in both classical and quantum reinforcement learning \cite{Williams1992}. It has been shown that QNN-based agents achieve comparable or even superior performance to classical deep networks on standard benchmarks \cite{Jerbi2021,Sequeira2023,Meyer2023}.

\textbf{Value-based methods.} The methods, such as Q-learning \cite{Sutton1998,Mnih2015}, employ QNNs to estimate the expected return of actions, i.e., the action-value function $Q(s,a)$. The optimal Q-function, denoted by $Q^*(s,a)$, satisfies the Bellman optimality equation,
\begin{equation}
Q^*(s,a)=\mathbb{E}_{s'\sim\mathcal{P}}\left[r(s,a)+\gamma\max_{a'}Q^*(s',a')\right],
\end{equation}
where $\mathcal{P}$ is the dynamics of the transition of the environment, $r$ is the immediate reward, and $\gamma$ denotes the discount factor. The advantage of QNN-based value estimation includes achieving comparable accuracy with significantly fewer parameters compared to classical NN \cite{Chen2020,Lokes2022,Chen2023,Chen2024,Lockwood2020,Skolik2022}. Further improvements in stability \cite{Liu2023} and generalization \cite{Periyasamy2024} have been obtained through variants such as double Q-learning and dueling architectures. Sensitivity of these methods to QNN hyperparameters has also been systematically reported \cite{Skolik2022,Getahun2023}.

\textbf{Actor-critic frameworks.} This type of strategy combines policy-based and value-based methods. In such a strategy, the actor, or equivalently the policy, is optimized using feedback from a learned critic, which is usually a value function. This reduces gradient variance while maintaining the flexibility of direct policy optimization \cite{Sutton1998}. Various QNN architectures have been explored for actor-critic implementations \cite{Kwak2021,Lan2021,Chen2023,Nagy2021}. These approaches typically use QNNs to represent the actor or critic, achieving performance comparable to classical counterparts, often with reduced model complexity \cite{Nagy2021}.

\subsubsection{Advanced Techniques}
Beyond the core algorithmic components, additional quantum mechanisms and properties can enhance the capability of QNNs for QRL.

Incorporating the long short-term memory of QRC as the agent has been shown to improve the efficiency of QRL even with reduced training overhead \cite{Chen2023c,Chen2024b}. The reservoir's intrinsic fading memory naturally captures temporal dependencies without backpropagation through time.

The introduction of QBM-based agents offers an energy-based alternative for policy function approximation. The policy is defined by the Boltzmann distribution,
\begin{equation}
\pi(a|s)=\frac{e^{-\beta E(s,a)}}{\sum_{a'}e^{-\beta E(s,a')}},
\end{equation}
where $E(s,a)$ is the energy of a QBM configuration. Initial studies demonstrated that such agents can outperform classical counterparts in discrete environments \cite{Crawford2018,Schenk2024}, and recent extensions show improved sample efficiency in hybrid actor–critic settings \cite{Schenk2024}. Deep QBM architectures further enable efficient learning in large action spaces, making such approaches viable on near-term devices \cite{Jerbi2021}.

Moreover, Grover-like quantum algorithms have been introduced to enhance the action selection process, aiming to accelerate decision-making in large action spaces through quadratic speedups in unstructured search \cite{Jerbi2021b,Teixeira2021,Sannia2023}. Other optimization techniques, including evolutionary algorithms \cite{Chen2022} and QNGD \cite{Meyer2023b}, have also shown promise for improving convergence rates and avoiding barren plateaus.

With these techniques, the architecture can also be extended to multi-agent settings, known as quantum multi-agent reinforcement learning. Advances over classical approaches have been reported in distributed settings, achieving higher rewards in edge computing and robotics tasks \cite{Yun2022,Yun2023}, or enabling fast adaptation with minimal data \cite{Yun2023b}.

\subsubsection{Implementations and Applications}
The study of noise impact on QNNs for QRL has revealed that performance is highly sensitive to both the type and level of noise \cite{Skolik2023b}. While some noise sources, such as decoherence, cause limited degradation, others, such as Gaussian perturbations, can be more detrimental. Interestingly, training with moderate noise may even improve robustness in certain cases, analogous to classical regularization effects. For a comprehensive review of QRL applications, we refer the reader to Refs. \cite{Meyer2022,Alomari2025}.

\subsection{Quantum Generative Adversarial Networks for QGL}

By employing one QNN as a generator for producing data that resembles real samples, and another QNN as a discriminator for evaluating the authenticity of the generated data, a quantum generative adversarial network (QGAN) can be constructed, as sketched in Fig. \ref{FIG:6}. This concept was introduced to enhance the capabilities of classical generative adversarial networks (GANs) by leveraging quantum effects such as superposition and entanglement. In classical machine learning, generative learning is an important strategy for generating data as supplements for insufficient training datasets, and GANs are among the most important routines for achieving this. The first work that proved the existence of a convergent point for QGANs was given by Lloyd and Weedbrook \cite{Lloyd2018}, which also provided an early investigation of the potential for efficient generation of complex data distributions. Of course, QGANs are not the only methods for QGL. For alternative strategies, one may refer to Ref. \cite{Tian2023}.

The typical characteristics of QGANs can be summarized as follows.

\subsubsection{Training Schemes}
\begin{figure}\centering
\includegraphics[width=.72\textwidth]{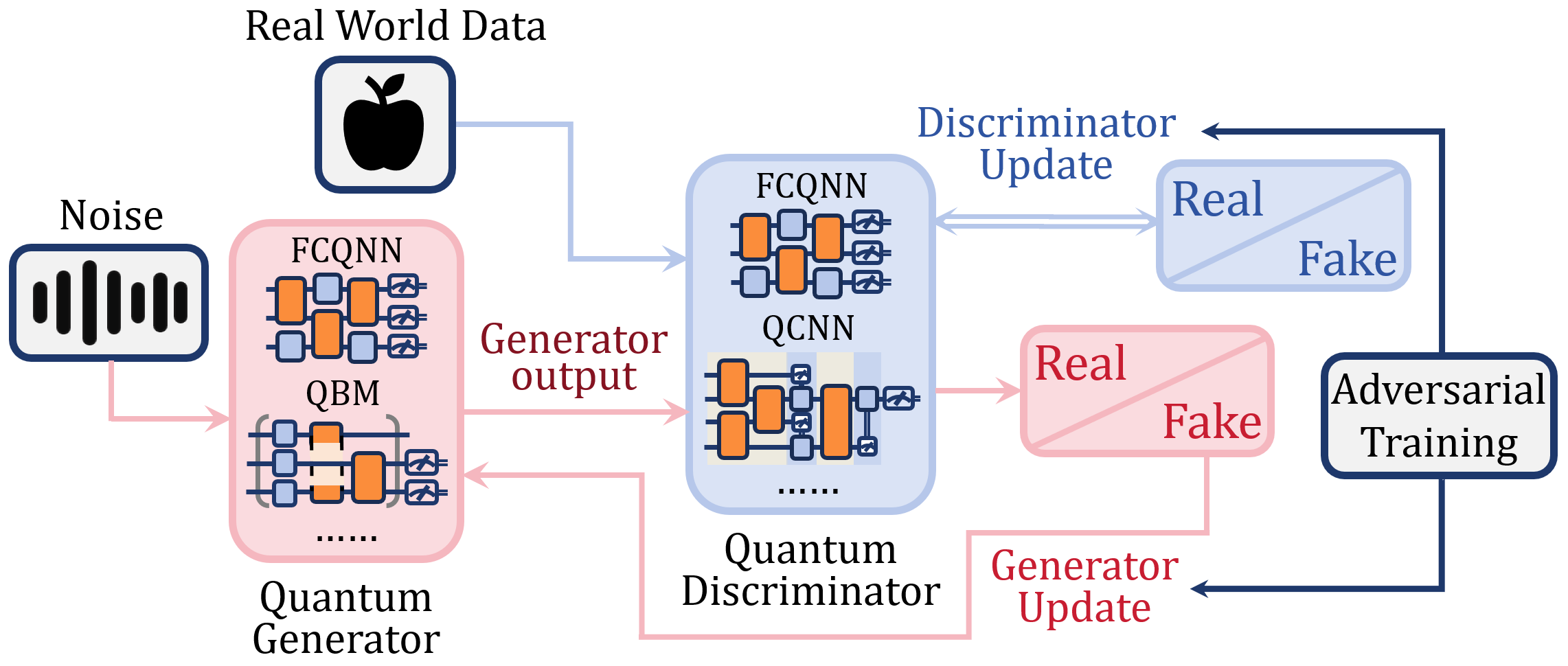}
\caption{A schematic of QGANs. The generator and discriminator are QNNs. The adversarial training of them finally reaches the point that the data output by the generator cannot be picked out from the real data by the discriminator.}\label{FIG:6}
\end{figure}

As briefly mentioned at the beginning of the Sec. \ref{Sec:CompQNN}, the training process of QGANs involves an adversarial game between the generator and the discriminator. More specifically, the generator tries to produce data indistinguishable from real data during training, while the discriminator attempts to correctly identify whether a given data point is real or generated, as shown in Fig.~\ref{FIG:6}. The objective function for a QGAN maximizes the probability that the generator successfully cheats the discriminator, i.e., the discriminator classifies fake data as real, while minimizing the probability that the discriminator correctly identifies fake data as generated. This adversarial process continues iteratively until the generator produces data with the same statistics as the real data, and the discriminator is unable to distinguish between them. This minimax objective can be expressed as
\begin{equation}
\min_{\theta_g}\max_{\theta_d}\mathcal{V}(\theta_g,\theta_d)=\mathbb{E}_{x\sim p_{\text{data}}(x)}[\log D_{\theta_d}(x)]+\mathbb{E}_{z\sim p_z(z)}[\log(1-D_{\theta_d}(G_{\theta_g}(z)))],
\end{equation}
where $G_{\theta_g}$ is the quantum generator circuit with parameters $\theta_g$, $D_{\theta_d}$ is the quantum discriminator with parameters $\theta_d$, $x$ represents real data samples, and $z$ denotes latent noise input \cite{Lloyd2018}. $p_{\text{data}}$ ($p_z$) denote the distribution of data (noise). For quantum data, where the true distribution is given by a density matrix $\sigma$, the objective simplifies to
\begin{equation}
\min_{\theta_g} \max_{T} \left( \text{Tr}[T\sigma] - \text{Tr}[T\rho(\theta_g)] \right),
\end{equation}
with $T$ being a positive operator-valued measurement element used by the discriminator and $\rho(\theta_g)$ the state produced by the generator \cite{Lloyd2018}.

The convergence point in the quantum case is proven to be unique and is easier to reach than in classical GANs, as the proof leverages the intrinsic probabilistic nature of quantum mechanics without requiring convexity assumptions \cite{Lloyd2018}. Recent advances have also shown that leveraging entangling operations between the generator output and true quantum data can further enhance the training process of QGANs, guaranteeing convergence to a Nash equilibrium under minimax optimization \cite{Niu2022}. This entangling QGAN overcomes mode collapse issues present in earlier proposals and demonstrates additional robustness against coherent errors \cite{Niu2022}.

\subsubsection{Performance and Applications}
QGANs have demonstrated several advantages over classical GANs. For instance, they can exhibit exponential speedup when dealing with high-dimensional data, as the quantum generator can efficiently represent distributions that would require exponentially many classical parameters \cite{Lloyd2018}. Additionally, quantum circuits can mitigate uncharacterized errors through adversarial training, making QGANs more robust to certain noise sources \cite{Niu2022}.

In practice, a fully quantum version of GAN, where both the generator and discriminator are quantum and the adversarial training is also performed by quantum algorithms, is not yet easy to implement on current quantum hardware due to limited qubit counts, coherence times, and the absence of efficient quantum optimizers. In practice, therefore, QGAN implementations typically adopt a hybrid approach. They employ QNNs only for the generator or the discriminator, while the optimization is performed by classical algorithms. This hybrid design significantly reduces the implementation complexity and hardware requirements. For instance, IonQ, in partnership with Zapata Computing, demonstrated the generation of high-resolution handwritten digits using such a hybrid quantum-classical GAN on trapped-ion quantum computers, achieving outstanding results with low error rates and high throughput \cite{IonQ2025Digits}. QGANs have also been used to load generic probability distributions for financial applications and to learn complex image datasets such as LSUN bedrooms for different circuit topologies \cite{Tian2023}. A notable variant that pushes toward more quantum capability is the quantum reservoir GAN, which employs QRC as the generator. This approach has been shown to achieve higher accuracy than both standard QGANs and classical neural networks on handwritten digit generation and CIFAR10 monochrome image generation tasks \cite{Wakaura2025}. 

Despite these advances, for such composite QNN architectures, it is not yet easy to observe clear quantum advantages over purely classical approaches on current state-of-the-art platforms due to hardware noise, limited qubit counts, and short coherence times. Consequently, hybrid quantum-classical implementations remain the most practical route for near-term applications, balancing quantum expressivity with classical optimization robustness. Research is actively ongoing to optimize circuit design, improve training stability, and scale QGANs to larger problem sizes as quantum hardware continues to mature \cite{Islam2025}.

\subsection{QNNs with Alignments for QTL}
Classical transfer learning is a well-established paradigm in machine learning that aims to leverage knowledge acquired from one task (usually called the source task) to improve learning performance on a related but distinct task (usually called the target task) \cite{Pan2010,Weiss2016}. Unlike traditional machine learning, which assumes that training and test data are drawn from the same distribution and that tasks are learned independently, transfer learning relaxes this assumption by recognizing that many real-world problems share underlying structures or features \cite{Pan2010}. For instance, a model trained to recognize objects in natural images (e.g., ImageNet) can be fine-tuned to detect tumors in medical scans with relatively few labeled examples \cite{Weiss2016}. The core challenge in transfer learning lies in determining what knowledge is transferable and how to adapt it effectively. This assessment is performed by an alignment subroutine, which measures the similarity between the source and target distributions and identifies which features or representations are suitable for reuse \cite{Pan2010}.

By introducing an alignment subroutine to QNNs, QTL can be achieved. Like QRL and QGL, QTL is the quantum counterpart of classical transfer learning, so that the core challenge is also the quantum alignment subroutine, determining what knowledge is transferable and how to adapt it effectively. 

For QNNs capable of handling QTL, as sketched in Fig. \ref{FIG:7}, knowledge transfer can occur in three forms \cite{Wang2023b}: (i) from a classical NN to a QNN, (ii) from a QNN to a classical NN, or (iii) between two QNNs. The alignment subroutine may be implemented as either a quantum algorithm (e.g., quantum kernel alignment (QKL) \cite{Havlíček2019}, quantum maximum mean discrepancy (QMMD) \cite{Kubler2019,Yao2026}) or a classical algorithm (e.g., principal component analysis \cite{Jolliffe2016}, autoencoder-based feature matching \cite{Costa2025}). However, most works focus on transferring knowledge from a classical NN to a QNN, which is relatively easier for practical applications due to the maturity of classical pre-trained models (e.g., ResNet, BERT, VGG) and the availability of large-scale pre-trained weights~\cite{Hu2025,MartinPerez2026}.

The general objective of QTL can be expressed as follows. Given a source task with data distribution $\mathcal{D}_S$ and a target task with distribution $\mathcal{D}_T$, where $\mathcal{D}_S$ and $\mathcal{D}_T$ are related but not identical, the goal is to leverage a pre-trained feature extractor $\phi_{\text{pre}}$ (classical or quantum) to improve learning on the target task. A QTL model typically takes the form
\begin{equation}
f_{\text{QTL}}(\mathbf{x})=h_{\theta_{\text{new}}}\bigl(\mathcal{E}\bigl(\phi_{\text{pre}}(\mathbf{x})\bigr)\bigr),
\end{equation}
where $\mathcal{E}$ is a quantum encoding map (e.g., angle embedding, amplitude embedding, etc), $\phi_{\text{pre}}$ is a fixed pre-trained feature extractor, and $h_{\theta_{\text{new}}}$ is a small trainable QNN adapted to the target task. The number of trainable parameters in $h_{\theta_{\text{new}}}$ is expected to be much smaller than the number of parameters in $\phi_{\text{pre}}$.

The alignment subroutine determines whether $\phi_{\text{pre}}(\mathbf{x})$ preserves sufficient relevant information for $\mathcal{D}_T$. By directly using the classical strategy of maximum mean discrepancy, a similar metric QMMD can be given by,
\begin{equation}
\mathrm{QMMD}^2(\mathcal{D}_S,\mathcal{D}_T)=\left\|\mathbb{E}_{x\sim\mathcal{D}_S}[\psi(x)]-\mathbb{E}_{y\sim\mathcal{D}_T}[\psi(y)]\right\|^2_{\mathcal{H}},
\end{equation}
where $\psi$ maps data to a reproducing kernel Hilbert space $\mathcal{H}$. 
A smaller QMMD indicates that the source and target distributions are closer, suggesting higher transferability. This technique also inspires the improvements on classical transfer techniques \cite{mccarty2025}. Another commonly used alignment metric for QTL is the kernel alignment score,
\begin{equation}
A(K_S,K_T)=\frac{\langle K_S,K_T\rangle_F}{\sqrt{\langle K_S,K_S\rangle_F\langle K_T,K_T\rangle_F}},
\end{equation}
where $\langle\cdot,\cdot\rangle_F$ denotes the Frobenius inner product. $K_S$ and $K_T$ are kernel matrices computed on source and target data, respectively \cite{Cristianini2002}. A score close to 1 indicates good alignment. In the quantum setting, these kernels can be replaced by quantum kernels of the form $K(x,x')=|\langle\phi(x)\phi(x')\rangle|^2$, where $|\phi(x)\rangle$ is a quantum feature map \cite{Havlíček2019}. This formulation allows quantum computers to compute kernel values, potentially offering quantum advantage in transfer learning tasks involving complex distribution alignments \cite{he2024}.

The investigations on QNNs for QTL are summarized below under core structures and practical applications.

\subsubsection{Two Ways of Categorizations}

The types of QNNs for QTL can be categorized along two complementary directions. One is categorized by knowledge type, and the other is categorized by classical involvement.

\textbf{Categorization by knowledge type.} The knowledge transferred can take several forms, each requiring different architectural choices.

\begin{itemize}
\item \textbf{Distribution-based transfer.} One transfers the \emph{distribution information} of a source dataset to aid training on a target dataset, provided the two datasets are closely distributed. In this case, a QNN may be employed as a generative model (e.g., QBM or QGAN) to produce samples approximating the source distribution. An alignment algorithm, such as QKL \cite{Havlíček2019} or QMMD \cite{Kubler2019,Yao2026}, then selects which generated samples are most suitable for training the target model. The transferability condition can be set to be proportional to value of $A(K_S,K_T)$ (sometimes only the value of $K$ \cite{Wang2021}), or proportional to the inverse of QMMD, or the combination of them both.

\item \textbf{Structure-based transfer.} Alternatively, one can transfer the \emph{structural information} of a pre-trained network to a QNN for modeling the target dataset. Typically, the pre-trained network is classical (e.g., ResNet-18 trained on ImageNet \cite{He2016ResNet}). Its output features are directly encoded into qubits via a quantum encoder $\mathcal{E}_{\text{en}}$. The alignment in such cases may be as simple as an identity map (i.e., directly feeding classical features into the QNN) or as complex as a trainable quantum encoder that learns to reweight or transform features for optimal quantum processing \cite{Havlíček2019,Wu2023}. The overall architecture is often called a quantum-classical hybrid NN for QTL \cite{Gokhale2020,Soto-Paredes2021,Acar2021,Mir2022,Azevedo2022,Huh2023,Dong2023,Alsharabi2023,Zhang2023,Decoodt2023,Buonaiuto2024,Ardeshir-Larijani2024,Dhara2024,Smaldone2024,Dey2024,Khatun2025}. The training objective for structure-based transfer is generally expressed by,
\begin{equation}
\mathcal{L}_{\text{QTL}}(\theta_{\text{new}})=\frac{1}{N_T}\sum_{i=1}^{N_T}\ell\bigl( h_{\theta_{\text{new}}}(\mathcal{E}_{\text{en}}(\phi_{\text{pre}}(x_i^{\text{target}}))), y_i^{\text{target}} \bigr),
\end{equation}
where only $\theta_{\text{new}}$ is optimized while $\phi_{\text{pre}}$ remains frozen.
$N_T$ denotes the number of target dataset samples. The function $\ell$ denotes the task-specific loss function (e.g., cross-entropy for classification or mean squared error for regression) that quantifies the discrepancy between the QNN prediction and the ground truth label, driving the optimization of the trainable parameters $\theta_{\text{new}}$ during target adaptation.

\item \textbf{Parameter-based transfer.} 
A less common but emerging approach involves transferring the actual weights or parameters of a pre-trained QNN to initialize another QNN for a related task. This is analogous to fine-tuning in classical deep learning. If the source QNN has parameters $\theta_{\text{pre}}$, the target QNN is initialized as $\theta_{\text{new}}^{(0)}=\theta_{\text{pre}}$ and then fine-tuned on the target task with a small learning rate. This approach requires that both QNNs share the same architecture, which is largely limited by current quantum technology.

Although direct QNN-to-QNN parameter transfer is challenging to implement, it has inspired an alternative and more practical direction. One can ues QNNs to generate fine-tuning parameters for classical NNs. Instead of transferring weights between two QNNs, these hybrid methods leverage quantum circuits to produce parameter-efficient adaptations for classical models. For instance, Quantum parameter adaptation methods use QNNs to generate fine-tuning parameters for classical models, decoupling the quantum resource requirement during inference and enabling scalable parameter-efficient learning \cite{Liu2025}. Similarly, Quantum-PEFT leverages Pauli parameterization to achieve logarithmic growth in trainable parameters, significantly outperforming classical low-rank adaptation methods in parameter efficiency while maintaining competitive performance \cite{KoikeAkino2025}. Besides, the quantum weighted tensor hybrid network further also demonstrates that quantum-enhanced fine-tuning can reduce trainable parameters by up to 76\% compared to LoRA while improving performance on certain benchmarks \cite{Kong2025}. 
\end{itemize}

\textbf{Categorization by classical involvement.} This categorization distinguishes methods based on whether a classical neural network is part of the architecture.

\begin{figure}\centering
\includegraphics[width=.72\textwidth]{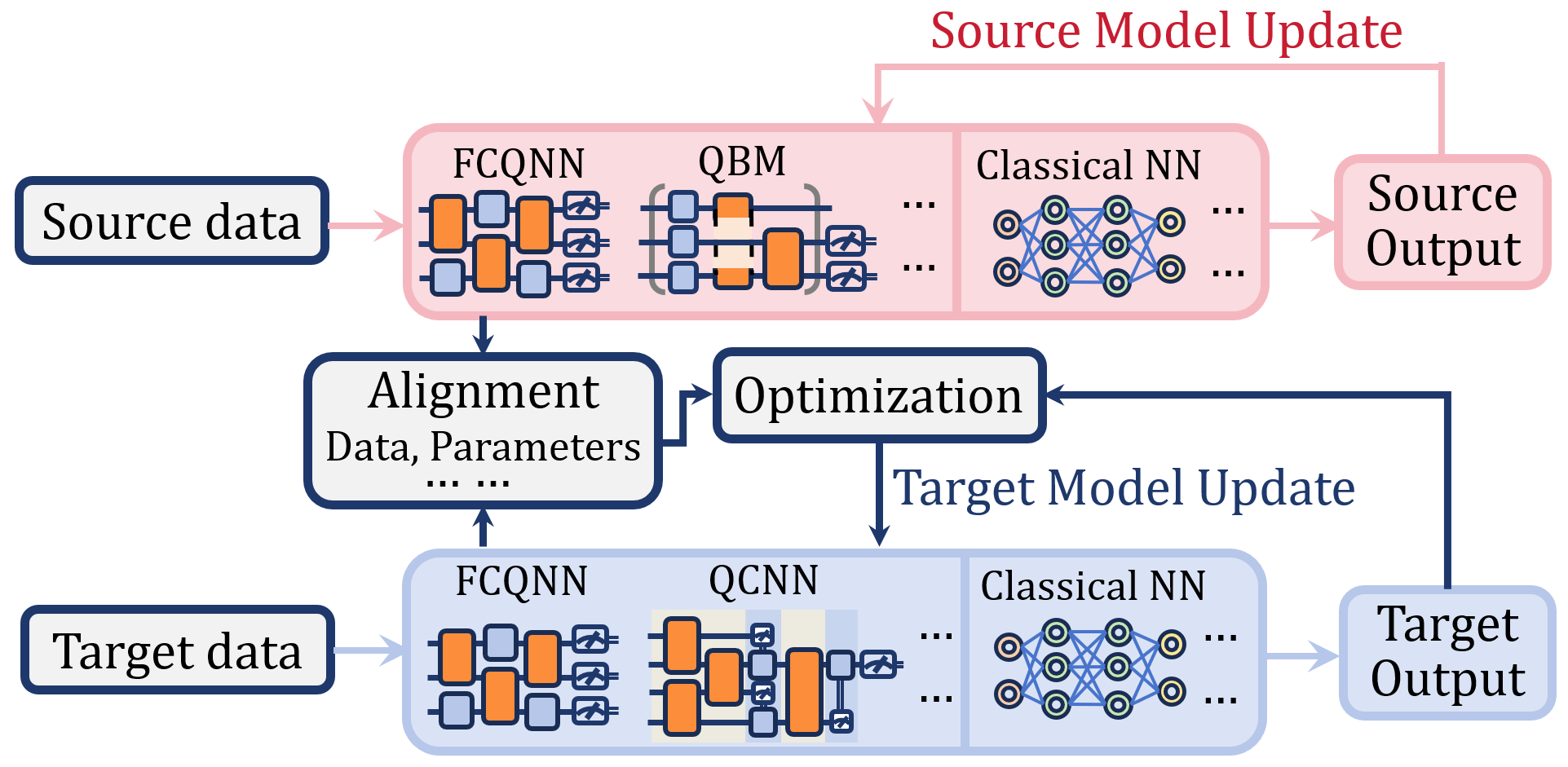}
\caption{A schematic of QNNs for transfer learning. The information learned from the source data are aligned to the learning process of target data.}\label{FIG:7}
\end{figure}

\begin{itemize}
\item \textbf{Pure quantum framework.} Works such as Refs. \cite{Wang2021,Wang2023} present QTL methods without classical NNs. Here, both the source and target feature extractors are quantum circuits. The alignment is performed using quantum kernel methods or QMMD computed directly on quantum states. This approach fully exploits quantum properties but requires sufficiently large and low-noise quantum hardware.

\item \textbf{Hybrid quantum-classical schemes.} Refs. \cite{Gokhale2020,Soto-Paredes2021,Acar2021,Mir2022,Azevedo2022,Huh2023,Dong2023,Alsharabi2023,Zhang2023,Decoodt2023,Buonaiuto2024,Ardeshir-Larijani2024,Dhara2024,Smaldone2024,Dey2024,Khatun2025,Liu2025,KoikeAkino2025,Kong2025} employ hybrid architectures involving both classical and quantum components. The hybrid approach remains dominant due to its ease of implementation on near-term devices. This pipeline is also illustrated in Fig.~\ref{FIG:7}.
\end{itemize}

\subsubsection{Practical Applications}

QNNs for QTL, espcially the hybrid quantum-classical schemes,  have been shown to be applicable to a wide variety of data types, including natural language, imaging, medical diagnostics, and industrial inspection.

\textbf{Natural language processing.} In natural language processing, it has shown that the hybrid schemes match or exceed the performance of state-of-the-art classical methods on several tasks. For acceptability judgments using the ItaCoLa dataset (Italian Corpus of Linguistic Acceptability), the scheme achieved comparable accuracy to fine-tuned BERT models while using significantly fewer parameters \cite{Buonaiuto2024}. For short text classification with SMS spam detection data, quantum-enhanced transfer learning achieved higher accuracy than classical baselines, particularly when training data was limited \cite{Ardeshir-Larijani2024}. The advantage stems from the quantum feature map's ability to capture non-linear correlations in text embeddings that classical linear classifiers miss.

\textbf{Image classification on standard benchmarks.} The hybrid schemes for QTL consistently outperform purely classical approaches on standard image benchmarks. On MNIST (handwritten digits), hybrid QTL achieves 98.5\% accuracy with only 4-8 qubits and fewer than 100 trainable parameters, compared to classical convolutional NNs requiring thousands of parameters \cite{Huh2023,Dhara2024}. On CIFAR-10 (natural images), QTL methods achieve competitive accuracy (above 85\%) using pre-trained ResNet features and a small QNN head, whereas training a classical convolutional NN from scratch on limited data typically underperforms \cite{Khatun2025}.

\textbf{Medical imaging and diagnostics.} QTL has shown particular promise in medical imaging, where labeled data is often scarce and expensive to obtain. For diabetic retinopathy detection from retinal fundus images, the hybrid schemes achieved higher sensitivity and specificity than classical transfer learning methods \cite{Mir2022}. For cardiomegaly detection in chest X-rays, QTL reduced false positives by 30\% compared to classical baselines \cite{Decoodt2023}. In knee osteoarthritis classification using X-ray images, QTL methods achieved state-of-the-art accuracy while requiring only 1/10 of the training data typically needed \cite{Dong2023}. These results suggest that QTL is particularly valuable in low-data regimes common in medical applications.

\textbf{Industrial and safety applications.} Beyond medical imaging, QTL has been successfully applied to COVID-19 protective mask recognition, where the system identifies whether a person is wearing a mask correctly, incorrectly, or not at all. QTL achieved 96\% accuracy with limited training data, outperforming classical methods trained on the same small dataset \cite{Soto-Paredes2021}. In industrial inspection tasks, QTL has been used for defect detection in manufacturing, leveraging pre-trained classical features from general object detection models and adapting them to specific defect types with minimal retraining \cite{Srinivasan2026}.

\textbf{Computational resource optimization.} Beyond accuracy improvements, QTL methods have demonstrated tangible reductions in training requirements. Instead of training a large QNN from scratch, which is computationally expensive and suffers from many limitations currently, QTL leverages a pre-trained classical NN as a fixed feature extractor. The classical NN, pre-trained on a large source dataset compresses raw high-dimensional data into a low-dimensional, semantically meaningful feature representation. A small QNN head is then trained on these compressed features to learn the target-specific mapping. Because the heavy feature extraction is already done by the parameter-fixed classical NN and only the small QNN head requires optimization, training times can be reduced by up to 80\% compared to training a QNN from scratch \cite{Acar2021}, and model parameter counts are often reduced by one to two orders of magnitude relative to fully classical models \cite{Zhang2023}. This makes QTL particularly attractive for near-term quantum devices, which have limited qubit counts, short coherence times, and are susceptible to barren plateaus.

Despite these successes, several challenges remain. The optimal choice of quantum encoding $\mathcal{E}_{\text{en}}$ for different feature types is not yet well understood. The alignment metric used to determine transferability often requires significant computational effort itself. Furthermore, the theoretical conditions under which QTL guarantees improvement over classical TL are still being explored. Research is actively ongoing to address these limitations and to scale QTL to larger problems as quantum hardware improves.

\section{QNNs with Other Structures}

Beyond the previously discussed families of QNNs, several other architectures incorporate specific structural innovations that enhance their expressivity, trainability, or applicability to particular tasks. Although these structures have received comparatively less attention in the literature, they show progress in the directions. A selection of representative examples is presented below.

\textbf{QNNs with randomness for enhanced expressivity.} A novel approach to boosting the expressivity of QNNs involves incorporating randomness into quantum circuits \cite{wu2024}. This strategy leverages random quantum operations to expand the representational capacity of the model, enabling it to capture more complex data distributions. Specifically, the authors introduce a random layer containing single-qubit gates sampled from a trainable ensemble pooling $\mathcal{E}=\{w_i, \hat{U}_{r,i}\}$ \cite{wu2024}. The QNN architecture consists of deterministic layers $\hat{U}_1$ and $\hat{U}_2$, with the random layer placed between them, followed by a classical function $f_{\boldsymbol{\beta}}$. For an input state $|\psi_m\rangle$, the final prediction is given by an ensemble average over a classical function of measurement outcomes,
\begin{equation}
\mathcal{P}_m=\sum_{i=1}^{N_r}w_i\mathcal{P}_{i,m}=\sum_{i=1}^{N_r}w_i f_{\boldsymbol{\beta}}(\mathbf{p}_{i,m}),
\end{equation}
where $w_i$ are the probabilities associated with each random unitary $\hat{U}_{r,i}$ in the ensemble, and $\mathbf{p}_{i,m}$ represents the measurement results from the quantum circuit. This formulation enables the model to accurately approximate arbitrary target operators using Uhlmann's theorem for majorization, which allows for observable learning.

\textbf{Quantum multilayer perceptron and the hopfield network based on it.} The quantum multilayer perceptron model \cite{shao2018} introduces a novel framework for implementing feed-forward neural networks on quantum computers. The core operation of a single quantum neuron in the $k$-th layer can be defined as a unitary transformation that maps an input quantum state to an output state encoding the activation function
\begin{equation}
|f(y^{(k)})\rangle=\bigotimes_{i=1}^{n_{k}}|f(y^{(k)}_i)\rangle,
\end{equation}
where $y^{(k)} = W^{(k)}x^{(k-1)}+b^{(k)}$ represents the weighted sum of inputs from the previous layer, $f(\cdot)$ is a nonlinear activation function implemented via quantum gates, and $n_k$ is the number of neurons in layer $k$ \cite{shao2018}. The model explicitly prepares the quantum state of the output signal in each layer using amplitude encoding and establishes a quantum learning algorithm for updating the weights $W^{(k)}$ based on gradient descent. This architecture achieves at least a quadratic speedup and an exponential speedup over classical algorithms when the matrix of weights has low rank \cite{shao2018}. As an extension, the paper also proposes an exponentially fast learning algorithm for the Hopfield network based on Hebbian learning rule, where patterns are stored in a superposition of attractor states, demonstrating the broad applicability of quantum methods to neural computation.

\textbf{Quantum autoencoder.} A quantum autoencoder (QAE) \cite{romero2017} is trained to compress a dataset of quantum states by discarding a latent register and then reconstructing the original state. The compression is achieved by a unitary $U(\boldsymbol{\theta})$ acting on an input state $|\psi_{\text{in}}\rangle$. The goal is to compress $|\psi_{\text{in}}\rangle$ into a smaller Hilbert space using a cost function based on a swap test between the input state and the reconstructed state. The cost function to be minimized is given by
\begin{equation}
C=1-\frac{1}{N}\sum_{i=1}^{N}\left|\langle\psi_{\text{in}}^{(i)}|V_{\text{ae}}(\boldsymbol{\theta})|0\rangle^{\otimes a}|\psi_{\text{in}}^{(i)}\rangle\right|^2,
\end{equation}
where $V_{\text{ae}}(\boldsymbol{\theta})$ is the unitary operation of the quantum autoencoder acting on the input state $|\psi_{\text{in}}^{(i)}\rangle$ and $a$ ancillary qubits initialized to $|0\rangle$. The cost function measures the overlap between the input state and the output of the autoencoder, which consists of the compressed state in the latent register and the original state in a reference register, achieved via a swap test. The QAE has been applied to compress ground states of the Hubbard model and molecular Hamiltonians, demonstrating substantial quantum data compression \cite{romero2017}.

\textbf{Quantum circuits with postselection for Bayesian learning.} 
Parameterized quantum circuits augmented with ancillary qubits and postselection exhibit strictly stronger expressive power than those without postselection \cite{du2020}. Consider a parameterized quantum circuit that generates the following state
\begin{equation}
\sum_{x}\sqrt{p(x)}|x\rangle|a(x)\rangle,
\end{equation}
where the first register encodes data $x$ and the ancilla register $|a(x)\rangle$ indicates some property of $x$. By postselecting on the ancilla being in the state $|0\rangle$, one obtains the conditional probability distribution
\begin{equation}
P(x|\text{postselect})=\frac{p(x)\mathbf{1}_{a(x)=0}}{\sum_{x'}p(x')\mathbf{1}_{a(x')=0}}.
\end{equation}
This technique is employed for Bayesian learning, enabling the model to learn prior probabilities rather than assuming them known \cite{du2020}. Such models are particularly advantageous in semisupervised learning scenarios where prior distributions are typically unknown. Numerical experiments on the Rigetti Forest platform demonstrate the performance of the proposed Bayesian quantum circuit \cite{du2020}.

\textbf{Hybrid quantum-classical NNs with neural tangent kernel.} The neural tangent kernel theory has been extended to a class of hybrid models consisting of a quantum data encoder followed by a classical NN \cite{nakaji2023}. In this framework, consider a quantum-classical NN composed of a quantum data encoder followed by a classical NN. The quantum part is randomly initialized according to unitary 2-designs, which serves as an effective feature extraction process for quantum states, while the classical part is randomly initialized according to Gaussian distributions \cite{nakaji2023}. In the neural tangent kernel regime where the number of nodes of the classical NN becomes infinitely large, the output of the entire hybrid quantum-classical NNs becomes a nonlinear function of the so called projected quantum kernel. The corresponding neural tangent kernel defined for the hybrid quantum-classical NNs is identical to the covariance matrix of a Gaussian process, which can be expressed as
\begin{equation}
K_{\text{qcNN}}(x,x')=\mathbb{E}_{\text{init}}\left[\nabla_{\boldsymbol{\theta}}f_{\text{qcNN}}(x)^\top \nabla_{\boldsymbol{\theta}}f_{\text{qcNN}}(x')\right],
\end{equation}
where $\boldsymbol{\theta}$ includes both classical and quantum parameters. $f_{\text{qcNN}}(x)$ denotes the final output of the hybrid quantum-classical NN, which is obtained by feeding the expectation values of quantum measurements into a classical NN. The expectation is taken over the random initialization. This hybrid architecture demonstrates a clear advantage over fully classical NNs and quantum NNs for the problem of learning quantum data-generating processes \cite{nakaji2023}.

\textbf{Hybrid quantum algorithm for molecular potential energy surfaces.} A hybrid quantum-classical algorithm employing a restricted Boltzmann machine has been developed to obtain accurate molecular potential energy surfaces \cite{xia2018}. The restricted Boltzmann machine encodes the electronic wavefunction as
\begin{equation}
P(\mathbf{v})=\frac{1}{Z}\sum_{\{h_j\}}e^{\sum_i a_i v_i+\sum_j b_j h_j+\sum_{i,j}w_{ij}v_i h_j},
\end{equation}
where $v_i\in\{-1, +1\}$ are the visible units representing the occupation of single-particle orbitals. $h_j\in\{-1, +1\}$ are the hidden units capturing correlations. $a_i$ and $b_j$ are trainable bias parameters for visible and hidden units respectively, $w_{ij}$ are the trainable connection weights between visible and hidden units, and $Z$ is the partition function ensuring proper normalization \cite{xia2018}. To address the sign problem inherent in electronic structure calculations (where wavefunction coefficients can be negative), the authors introduce a three-layer restricted Boltzmann machine architecture that adds a separate sign function
\begin{equation}
\phi(\mathbf{v})=\sqrt{P(\mathbf{v})}\cdot s(\mathbf{v}),~\text{with}~s(\mathbf{v})=\tanh\left(\sum_i d_i v_i+c\right),
\end{equation}
where $d_i$ and $c$ are additional trainable parameters for the sign layer \cite{xia2018}. This hybrid quantum algorithm optimizes the restricted Boltzmann machine parameters by exploiting a quantum subroutine to help optimize the underlying objective function, efficiently computing the electronic ground state energy for small molecular systems such as H$_2$, LiH, and H$_2$O at specific locations on their potential energy surfaces \cite{xia2018}.

Beyond the examples listed above, many other QNN structures continue to emerge. The integration of diverse architectural motifs, such as randomness, postselection, autoencoding, kernel methods, and hybrid designs, has significantly expanded the capabilities and applicability of QNNs. As quantum computing technology advances, these innovative structures are expected to play a crucial role in unlocking the full potential of quantum machine learning.

\section{Quantum Machine Learning without NNs}

While QNNs have garnered significant attention in the field of quantum machine learning, there are numerous other quantum algorithms and techniques that offer promising alternatives for various machine learning tasks without relying on NN structures. These methods can be broadly categorized into several key areas, each leveraging distinct quantum computational advantages.

\textbf{Quantum support vector machines.} Quantum support vector machines (QSVMs) are a powerful alternative to classical SVMs, leveraging quantum computing to enhance classification tasks \cite{Rebentrost2014,Havlíček2019}. The core idea of a classical SVM is to find a hyperplane that separates data points of different classes by maximizing the margin
\begin{equation}
\min_{\mathbf{w},b}\frac{1}{2}\|\mathbf{w}\|^2~~~~\text{subject to}~~y_i(\langle\mathbf{w},\mathbf{x}_i\rangle+b)\geq 1,
\end{equation}
where $\mathbf{w}$ is the normal vector to the hyperplane, $b$ is the bias term, and $(\mathbf{x}_i,y_i)$ are training samples with labels $y_i\in\{-1,+1\}$ \cite{cortes1995}. The dual formulation of this optimization problem reveals the kernel trick
\begin{equation}
\max_{\boldsymbol{\alpha}}\sum_{i=1}^{M}\alpha_i-\frac{1}{2}\sum_{i,j=1}^{M}\alpha_i\alpha_j y_i y_j K(\mathbf{x}_i,\mathbf{x}_j)~~~~\text{subject to}~~\alpha_i\geq 0,~\sum_{i=1}^{M}\alpha_i y_i=0,
\end{equation}
where $K(\mathbf{x}_i,\mathbf{x}_j)=\langle\varphi(\mathbf{x}_i),\varphi(\mathbf{x}_j)\rangle$ is the kernel function and $\varphi$ maps data into a high-dimensional feature space \cite{boser1992}. QSVMs utilize quantum circuits to evaluate kernel functions that are classically intractable, by encoding data into quantum states via a feature map circuit $U_\varphi(\mathbf{x})$
\begin{equation}
\kappa(\mathbf{x}_i,\mathbf{x}_j)=\left|\langle 0|^{\otimes n}U_\varphi^\dagger(\mathbf{x}_j) U_\varphi(\mathbf{x}_i)|0\rangle^{\otimes n}\right|^2,
\end{equation}
which corresponds to the fidelity between two quantum states encoding the data points \cite{Havlíček2019,schuld2020}. This approach can handle high-dimensional data more efficiently and potentially achieves exponential speedup in kernel evaluation for certain feature maps \cite{Rebentrost2014,liu2021b}.

\textbf{Quantum kernel methods.} Quantum kernel methods represent another significant advancement in quantum machine learning, extending the classical kernel framework to quantum feature spaces \cite{Schuld2019,schuld2020}. Unlike QSVMs which specifically target classification, quantum kernel methods encompass a broader class of algorithms including kernel ridge regression, kernel principal component analysis, and clustering \cite{Schnabel2024}. Unlike QNNs where parameters are trained directly on the quantum device, quantum kernel methods treat the quantum feature map as fixed ones. Learning occurs entirely in the classical domain using the kernel values computed by the quantum computer. Formally, a quantum kernel is defined via a parameterized quantum circuit that maps classical data to quantum states
\begin{equation}
|\psi(\mathbf{x})\rangle=\mathcal{U}_{\text{en}}(\mathbf{x})|0\rangle^{\otimes n},
\end{equation}
where $\mathcal{U}_{\text{en}}(\mathbf{x})$ is a data-encoding unitary circuit \cite{Schuld2019}. The kernel function is then given by the inner product between quantum states
\begin{equation}
K(\mathbf{x}_i,\mathbf{x}_j)=|\langle\psi(\mathbf{x}_i)|\psi(\mathbf{x}_j)\rangle|^2,
\end{equation}
which corresponds to the geometric inner product of the encoded quantum states \cite{Schuld2019}. For an $n$-qubit system, the quantum feature space has dimension $2^n$, enabling exponential expansion of the feature space with only linear growth in qubits \cite{Havlíček2019}. Two primary families of quantum kernels have been studied. One is the fidelity quantum kernels (FQKs) and projected quantum kernels (PQKs) \cite{Schnabel2024}. FQKs directly measure the overlap between quantum states via the swap test, while PQKs leverage reduced density matrices to construct kernels that are invariant under certain symmetries \cite{Schnabel2024,huang2021b}. A work has highlighted the critical importance of hyperparameter tuning, particularly kernel bandwidth selection, for achieving optimal performance in quantum kernel methods \cite{Shaydulin2022}. These methods enable the processing of complex data structures more efficiently than classical methods, with theoretical guarantees of quantum advantage for specific learning tasks \cite{huang2021b,liu2021b}.

\textbf{Quantum principal component analysis.} Quantum principal component analysis (QPCA) aims to identify the principal components of large datasets using quantum algorithms, offering significant speedup compared to classical counterparts \cite{Lloyd2014,nghiem2025}. The classical PCA problem involves diagonalizing the covariance matrix $C=\frac{1}{M}\sum_{i=1}^{M}\mathbf{x}_i\mathbf{x}_i^T$ to extract its largest eigenvalues and corresponding eigenvectors \cite{pearson1901}. The quantum algorithm, proposed by Lloyd, Mohseni, and Rebentrost, assumes the ability to prepare the density matrix $\rho=\sum_i p_i |\psi_i\rangle\langle\psi_i|$, which corresponds to the covariance matrix for centered datasets \cite{Lloyd2014}. Using the technique of density matrix exponentiation, QPCA can reveal the principal components via the repeated application of swap operations
\begin{equation}
\operatorname{Tr}_P e^{-iS\Delta t} \rho \otimes \sigma e^{iS\Delta t} = \sigma - i\Delta t[\rho, \sigma] + O(\Delta t^2),
\end{equation}
where $S$ is the swap operator acting on two registers, $\operatorname{Tr}_P$ denotes the partial trace over the first register, and $\Delta t$ is a small time increment. This algorithm achieves an exponential speedup in the dimension of the feature space compared to classical PCA under certain conditions, with complexity $O(\log d)$ rather than $O(d)$ for $d$-dimensional data \cite{Lloyd2014}. Recent advances have provided improved protocols for preparing the covariance matrix on quantum computers, including methods for uncentered datasets and connections to the quantum singular value transformation framework \cite{nghiem2025,gordon2022}. QPCA has been demonstrated on both synthetic and real-world datasets, including the MNIST handwritten digit dataset and molecular ground-state datasets, showing potential for quantum advantage in dimensionality reduction tasks \cite{gordon2022,lin2024}.

\textbf{Other quantum classifiers.} Various other quantum classifiers are being explored, each tailored to exploit the unique properties of quantum mechanics. These include quantum decision trees that utilize conditional probabilities and information gain to classify quantum states \cite{li2025}, quantum nearest-neighbor algorithms that compute distance metrics such as inner products and Euclidean distances with polynomial, exponential, or even super-exponential query reductions \cite{wiebe2014b}, quantum boosting and ensemble methods such as QAdaBoost and QRealBoost that combine weak quantum learners to construct stronger ones with provable quadratic speedups \cite{wang2024v,chatterjee2022}, and variational quantum classifiers that differ from standard QNN architectures in their ansatz design \cite{Havlíček2019,schuld2020}. For a detailed review of these emerging approaches, the reader is referred to Ref. \cite{Li2022,benedetti2019}.

\section{The Problem of Barren Plateaus and Relevant Progress}

Despite the progress in the structures of QNNs and their performance, a major obstacle in the advancement of QNNs is the problem of barren plateaus. This phenomenon causes the loss function landscape to become exponentially flat as the number of qubits increases, rendering training exceedingly difficult. This section examines the definition, origins, and consequences of barren plateaus. Meanwhile, recent progress is also briefly reviewed, including their intimate connection to classical simulability.

\subsection{Formal Definition}
Consider a parameterized quantum circuit $U(\boldsymbol{\theta})$ acting on an initial state $\rho$, with a measurement observable $O$. The loss function (also the cost function in some researches) takes the fundamental form
\begin{equation}\label{eq:lossTr}
\ell(\boldsymbol{\theta})=\operatorname{Tr}\left[U(\boldsymbol{\theta})\rho U^{\dagger}(\boldsymbol{\theta})O\right],
\end{equation}
where the dependence on $\rho$ and $O$ is implicit. A barren plateau is defined as the exponential concentration of the loss function landscape around its mean as the number of qubits $n$ increases. Specifically, if the variance of $\ell(\boldsymbol{\theta})$ decays exponentially with $n$,
\begin{equation}
\operatorname{Var}_{\boldsymbol{\theta}\sim\mathcal{P}}\left[\ell(\boldsymbol{\theta})\right] \lesssim \frac{1}{b^n},~~b>1,
\end{equation}
then a barren plateau exists. $\boldsymbol{\theta}\sim\mathcal{P}$ means $\boldsymbol{\theta}$ are sampled from the probability distribution $\mathcal{P}$. Conversely, the absence of barren plateaus is characterized by a variance that decays at most polynomially,
\begin{equation}
\operatorname{Var}_{\boldsymbol{\theta}\sim\mathcal{P}}\left[\ell(\boldsymbol{\theta})\right]\gtrsim \frac{1}{\operatorname{poly}(n)}.
\end{equation}
By Chebyshev's inequality, the probability that the loss function deviates from its mean by more than $\delta$ is exponentially small,
\begin{equation}
\operatorname{Pr}_{\boldsymbol{\theta}\sim\mathcal{P}}\left(|\ell(\boldsymbol{\theta})-\mathbb{E}[\ell(\boldsymbol{\theta})]|\geq\delta\right)\lesssim\frac{1}{b^n}.
\end{equation}
One can also define a deterministic barren plateau, where the loss landscape is globally flat
\begin{equation}
|\ell(\boldsymbol{\theta})-\ell_0|\lesssim\frac{1}{b^n},~~|\partial_{\theta} \ell(\boldsymbol{\theta})|\lesssim\frac{1}{b^n},
\end{equation}
meaning that no effective descent direction exists \cite{McClean2018}.

It is important to note that strategies such as using higher-order gradients, gradient-free optimizers, QNGD, or noise mitigation techniques cannot resolve barren plateaus, as these approaches do not fundamentally alter the geometric properties of the landscape \cite{arrasmith2021}.

Another important notice is that the barren plateau phenomenon in QNNs {\it differs fundamentally} from gradient vanishing in classical deep networks. In the quantum case, the Hilbert space dimension grows exponentially with the number of qubits, causing the evolved state to become uniformly distributed in a high-dimensional space, leading to an exponentially small inner product with the measurement operator. This is a geometric property of the quantum system, which involves factors such as the circuit depth, ansatz, observables, loss functions, etc. In contrast, classical gradient vanishing arises from the chain rule in deep networks, where products of activation function derivatives and weights accumulate to produce exponential decay with depth. Classical techniques such as normalization and residual connections cannot be directly applied to quantum circuits due to the no-cloning theorem.

\subsection{Proven Cases of Barren Plateaus}

Building on the definitions above, rigorous proofs have demonstrated that barren plateaus necessarily arise in the following specific cases.

\textbf{Haar measure and unitary $t$-designs.} To characterize the randomness of parameterized quantum circuits, we consider the unitary group $U(d)$ of dimension $d=2^n$. The Haar measure $d\mu_{\text{Haar}}(U)\equiv dU$ satisfies left and right invariance,
\begin{equation}
\forall U_0\in U_d, ~~~\int_{U_d}f(U_0 U)dU=\int_{U_d}f(U)dU=\int_{U_d}f(U U_0)dU,
\end{equation}
with normalization $\int_{U_d}dU=1$ \cite{McClean2018}. A unitary $t$-design is a finite set of unitaries that approximates the Haar measure up to the $t$-th moment, shown by
\begin{equation}
\sum_{m=1}^{M}p_m U_m^{\otimes t}X(U_m^{\dagger})^{\otimes t}=\int_{U(d)}U^{\otimes t} X(U^{\dagger})^{\otimes t}dU.
\end{equation}
For $t=1$, the first moment condition is
\begin{equation}
\sum_{m=1}^{M} p_m U_m X U_m^{\dagger}=\int_{U(d)}UXU^{\dagger}dU=\frac{\operatorname{Tr}(X)}{d}I.
\end{equation}
In QNN research, the loss function involves first moments, while the variance involves second moments. A circuit that forms a $2$-design is considered sufficiently random to induce barren plateaus.

\textbf{Deep circuits induce barren plateaus.} Consider a QNN structure of depth $L$
\begin{equation}
U(\boldsymbol{\theta})=\prod_{l=1}^{L} U_l(\boldsymbol{\theta}_l)W_l,
\end{equation}
where $U_l(\boldsymbol{\theta}_l)=\exp(-i\theta_l V_l)$ with $V_l$ Hermitian, and $W_l$ are fixed entangling unitaries \cite{McClean2018}. Let the initial state be $|0\rangle$ and the observable be $H$. The partial derivative of the loss function with respect to parameter $\theta_k$ is
\begin{equation}
\partial_k \ell=i\langle 0|U^{\dagger}[V_k,U_k^{\dagger}HU_+]U_-|0\rangle,
\end{equation}
where $U_-$ and $U_+$ denote the circuit parts before and after the $k$-th gate, respectively. Assuming $U_-$ forms a $1$-design, we obtain,
\begin{equation}
\sum_m p_m^- U_m\rho U_m^{\dagger}=\frac{I}{d}, \quad d=2^n.
\end{equation}
The expectation of the gradient then vanishes,
\begin{equation}
\mathbb{E}[\partial_k\ell]=\frac{1}{d}\sum_{m'}p_{m'}^+\operatorname{Tr}\left([V_k, U_{m'}^{\dagger}H U_{m'}]\right)=0,
\end{equation}
since the trace of a commutator is always zero. The variance scales as
\begin{equation}
\operatorname{Var}(\partial_k\ell)\approx-\frac{1}{d^2-1} \operatorname{Tr}\left(\sum_{m'}p_{m'}A_{m'}^2\right),
\end{equation}
which decays exponentially with the number of qubits because $d=2^n$ \cite{McClean2018}.

\textbf{Global observables induce barren plateaus in shallow circuits.} Even for shallow circuits with depth $L\sim\operatorname{poly}(\log n)$, barren plateaus can arise if the measurement observable is global. A global observable acts nontrivially on all qubits, such as
\begin{equation}
O_{\text{global}}=|0\rangle\langle 0|^{\otimes n},
\end{equation}
which yields a nonzero measurement only when the quantum state is exactly $|00\ldots 0\rangle$. For a random quantum state, the probability of this event is exponentially small. In contrast, a local observable acts only on a few qubits
\begin{equation}
O_{\text{local}}=|0\rangle\langle 0|\otimes I_2\otimes I_3\otimes\cdots\otimes I_n.
\end{equation}
For hardware-efficient ansatze where each two-qubit gate forms a $2$-design, it has been shown that global observables lead to exponentially decaying gradient variances, while local observables yield polynomially decaying variances, thus avoiding barren plateaus \cite{cerezo2021}.

\subsection{Absence of Barren Plateaus vs. Classical Simulability}
As briefly mentioned in Sec. \ref{ssec:AdvQCNN}, QCNNs have been shown to avoid barren plateaus under specific conditions \cite{Pesah2021}. The absence arises from two key features: the circuit is shallow (depth logarithmic in the number of qubits) and the measurement observables are local. The proof methodology is similar to that for hardware-efficient circuits with local cost functions, though the specific architectural details of QCNNs lead to different computational constants.

A particularly interesting finding, given the absence of barren plateaus in QCNNs, is that a recent work has demonstrated these architectures are able to effectively simulate by classical computing \cite{Bermejo2026,angrisani2025}. The core idea is to expand the observable in the Pauli basis
\begin{equation}
O=\sum_{\alpha=1}^{4^n}c_{\alpha}P_{\alpha},~~P_{\alpha}=P_1\otimes P_2\otimes\cdots\otimes P_n,~~P_i\in\{I,X,Y,Z\},
\end{equation}
and then truncate the expansion to low-weight Pauli operators. The weight of a Pauli operator is defined as the number of non-identity terms in the tensor product. In randomly initialized QCNNs, the contribution of high-weight Pauli operators decays exponentially with weight \cite{Bermejo2026,angrisani2025}. Therefore, only low-weight operators need to be retained.

Simulation proceeds in the Heisenberg picture, which is based on the loss function of the same form as Eq. (\ref{eq:lossTr}). The classical simulation algorithm can be performed as follows,
\begin{enumerate}
    \item Expand $O$ in the Pauli basis, $O=\sum_{\alpha}c_{\alpha}P_{\alpha}$.
    \item Perform low-weight truncation to obtain $O_{\text{low}}=\sum_{\alpha=1}^{t}c_{\alpha} P_{\alpha}$ with $t\ll 4^n$.
    \item Propagate Pauli operators layer by layer, $U_L^{\dagger} O_{\text{low}} U_L$, truncating high-weight operators after each layer.
    \item Compute the expectation value $\operatorname{Tr}[\rho O_{\text{final}}]$.
\end{enumerate}
If the input state $\rho$ is classical (e.g., a matrix product state), the entire computation can be performed classically without a quantum computer \cite{Bermejo2026,angrisani2025}. A similar strategy has also been considered in optical computing literature \cite{li2024s}.

The classical simulability of QCNNs raises a profound question: does the absence of barren plateaus necessarily imply that a quantum model is classically simulable? Recent theoretical work has explored this exact connection \cite{cerezo2025}. The key insight comes from an operator space perspective. The loss function can be viewed as an inner product in the operator space of dimension $4^n$
\begin{equation}
\ell(\boldsymbol{\theta})=\langle\rho,U^{\dagger}(\boldsymbol{\theta})OU(\boldsymbol{\theta})\rangle.
\end{equation}
When both $\rho$ and $U^{\dagger}(\boldsymbol{\theta})OU(\boldsymbol{\theta})$ can be restricted to a polynomially sized subspace (e.g., the space of low-weight Pauli operators), two consequences follow simultaneously: (1) the variance of $\ell(\boldsymbol{\theta})$ decays at most polynomially (i.e., no barren plateau), and (2) the entire computation can be performed classically by tracking only the polynomially many basis elements. Conversely, if the dynamics escape this low-weight subspace, the variance tends to decay exponentially (i.e., barren plateau emerges), but quantum advantage may become possible \cite{cerezo2025}.

This suggests a potential trade-off, i.e., provable absence of barren plateaus may imply classical simulability, at least for natural families of variational quantum models. However, this is not an absolute theorem. Exceptions exist for artificially constructed circuits (e.g., those mimicking Shor's algorithm), but such constructions lack practical utility for quantum machine learning. Therefore, while avoiding barren plateaus is necessary for trainability, it may not be sufficient for achieving quantum advantage. Indeed, it might actually indicate that the model is classically simulable \cite{Bermejo2026,cerezo2025}. This tension between trainability and expressivity remains a central open question in the field.

\subsection{The Additional Challenge of Local Minima}
Even when barren plateaus are absent, QNNs face another critical challenge, which is the proliferation of poor local minima. For shallow or local circuits that avoid barren plateaus, it has been shown that the loss function landscape converges to Wishart hyperspherical random fields in the underparameterized regime. In this regime, only a superpolynomially small fraction of local minima are close to the global optimum. The vast majority are poor-quality local minima, and gradient-based algorithms readily become trapped \cite{anschuetz2022}. 

Moreover, barren plateaus themselves are accompanied by an exponential number of poor local minima. The total loss function can be decomposed as $\ell(\boldsymbol{\theta})=\sum_i \ell_i(\boldsymbol{\theta})$. If each $\ell_i(\boldsymbol{\theta})$ exhibits a barren plateau, then a configuration that minimizes one term while leaving others flat becomes a local minimum of the total cost. Strategies designed to avoid barren plateaus, such as near-identity initialization or near-Clifford initialization, produce large gradients but place the initial parameters near Clifford points, which are themselves surrounded by poor local minima, leading to immediate trapping \cite{nemkov2025}.

\section{Discussions and Conclusions}

Despite the significant progress made in recent years, the field of QNNs and quantum machine learning still faces numerous challenges. These include addressing the limitations of current quantum hardware, improving noise resilience, and developing more efficient training algorithms. Additionally, the theoretical understanding of QNNs, particularly in terms of their expressivity, generalization capabilities, and scalability, remains an area of active research. In light of these considerations, several potential topics that may prove interesting for future research are listed below.

\textbf{Quantum transformer models.} In parallel to the classical success of transformer architectures, there are ongoing efforts to develop quantum counterparts. Although classical transformers are well established, significant difficulties remain in designing a quantum version of the attention mechanism, which is one of the key components in classical transformers. The challenge lies in implementing the quadratic pairwise interactions required for attention using quantum operations that respect linearity and unitarity. There is potential for a novel attention mechanism that leverages the unique aspects of quantum dynamics, such as superposition and interference.

One notable example is Quixer, a quantum transformer model that uses linear combinations of unitaries and the quantum singular value transform as building blocks \cite{Khatri2024}. Quixer prepares a superposition of tokens and applies a trainable non-linear transformation through quantum circuits. Initial results on language modeling tasks show performance competitive with classical models of comparable size.

Subsequent work has expanded the quantum transformer landscape in several directions. FitQT introduces a fully integrated and trainable quantum transformer that unifies three essential components, which includes a quantum multi-head attention mechanism employing a scaled Gaussian kernel, a quantum feed-forward network, and quantum residual connections to stabilize gradient flow \cite{Pham2026}. Evaluations on the IMDb dataset demonstrate that FitQT improves accuracy by 14.89\% over baseline quantum self-attention networks while requiring only 2,595 parameters. Another approach, called quantum adaptive self attention (QASA), adopts a principle of architectural parsimony, replacing only the value projection in a single encoder layer with a parameterized quantum circuit using just 36 trainable quantum parameters \cite{Chen2025}. QASA achieves a 6.0\% MAE reduction on the real-world ETTh1 dataset and outperforms models with 2 to 4 times more quantum parameters, suggesting that maximal quantum benefit comes not from maximizing quantum resources but from strategically placing minimal quantum computation.

Beyond language tasks, quantum transformers have been applied to diverse domains. QuantumSMoE combines a vision transformer with a mixture-of-experts layer for high-fidelity surface code decoding in quantum error correction, outperforming both classical baselines such as Minimum Weight Perfect Matching and state-of-the-art ML-based decoders on the toric code \cite{Viet2026}. A quantum-enhanced vision transformer for flood detection using remote sensing imagery achieves a significant accuracy improvement from 84.48\% to 94.47\% compared to classical ViT baselines \cite{Maity2026}. The QSTAformer embeds parameterized quantum circuits into attention mechanisms for short-term voltage stability assessment in power systems, demonstrating competitive accuracy and stronger robustness against adversarial attacks \cite{Li2025Q}. Other investigations, such as proposals for quantum vision transformers, have also been reported, extending the transformer paradigm to image recognition tasks on quantum devices \cite{Cherrat2024}. These efforts, while still in early stages, suggest that quantum attention mechanisms may offer unique advantages not replicable by classical transformers.

\textbf{Could Barren plateaus be helpful?} A counterintuitive emerging perspective is that barren plateaus could be harnessed as a useful feature rather than a bug. To appreciate this possibility, we first review the joint-embedding predictive architecture (JEPA), a self-supervised learning framework introduced by LeCun \cite{lecun2022J}. Unlike generative approaches that reconstruct inputs in pixel space (e.g., autoencoders or diffusion models) or contrastive methods that require negative samples, JEPA learns by predicting the latent representation of one part of an input from the representation of another part \cite{assran2023iJ}. Specifically, given an input $\mathbf{x}$, JEPA defines two views, a context $c$ and a target $t$. The context encoder $f_\theta$ maps the context to an embedding $f_\theta(c)$, while the target encoder $g_\phi$ maps the target to an embedding $g_\phi(t)$. A predictor $p_\psi$ is then trained to predict the target embedding from the context embedding
\begin{equation}
\mathcal{L}_{\text{JEPA}}=\mathbb{E}_{\mathbf{x}}\left[\|p_\psi(f_\theta(c))-g_\phi(t) \|_2^2\right],
\end{equation}
where $\|\cdot\|_2$ is the Euclidean norm ($L_2$ norm). Crucially, to prevent representation collapse, in which case the encoder outputs a constant embedding for all inputs, JEPA relies on explicit regularization techniques such as variance regularization, stop-gradient operations, or exponential moving average updates of the target encoder \cite{bardes2022V,grill2020B}. These heuristics, while effective, add complexity and hyperparameter sensitivity to training.

The central challenge JEPA addresses is the following. Without careful regularization, the predictor can simply learn to output a constant embedding regardless of the context, trivially satisfying the prediction loss while learning no useful representation of the data \cite{assran2023iJ}. This collapse problem is particularly acute when the target embeddings lack sufficient diversity or when the predictor is too expressive relative to the encoder.

What JEPA fundamentally requires to avoid collapse is a representational space where distinct inputs map to sufficiently separated embeddings, a property we call \textit{automatic representational diversity}. Classical JEPA implementations enforce this diversity through explicit regularization terms, such as variance regularization  and covariance regularization \cite{bardes2022V}, or through architectural heuristics like stop-gradient operations and exponential moving average target encoders \cite{grill2020B}.

Remarkably, this required property seems to emerge naturally in QNNs that exhibit barren plateaus. For a QNN whose parameterized circuit forms a $2$-design, the output quantum states $|\psi(\mathbf{x})\rangle$ for distinct inputs are distributed approximately Haar-uniformly on the unit sphere in Hilbert space \cite{McClean2018}. For Haar-random quantum states, the expected squared inner product between two independently drawn states is exponentially small
\begin{equation}
\mathbb{E}_{|\psi\rangle,|\phi\rangle\sim\text{Haar}}\left[|\langle\psi|\phi\rangle|^2 \right]=\frac{1}{d}=\frac{1}{2^n}.
\end{equation}
This means the embedding space is exponentially large, and random embeddings are nearly orthogonal with high probability. A predictor that attempts to output a constant embedding would incur an exponentially large prediction loss.



However, a subtle challenge also arises. While Haar-random states are nearly orthogonal, which prevents collapse. This also means that distinguishing them requires an exponential number of measurements in the worst case. This raises a critical question: if the target quantum embeddings are Haar-random: can a predictor learn to accurately predict them using only polynomially many measurement shots? The answer, unfortunately, is no for arbitrary Haar-random states. The exponential distinguishability barrier seems to undermine the very feasibility of quantum JEPA.

Nevertheless, this challenge does not necessarily invalidate the proposal. The key insight is that the target embeddings in a practical quantum JEPA would not be arbitrary Haar-random states. Rather, they would be structured states generated by a parameterized encoder that evolves slowly during training. Under this more realistic setting, the distinguishability barrier may be surmountable. Specifically, the proposal that barren plateaus could serve as an anti-collapse mechanism for quantum JEPA remains \textit{speculative but potentially viable} under the following conditions. First, the predictor need not output the exact target quantum state. It suffices to produce a classical embedding that correlates with measurement outcomes of the target state. Second, the target encoder can be updated slowly (e.g., via exponential moving average), allowing the predictor to track a smoothly evolving latent space rather than confronting arbitrary Haar-random targets at each step. Third, recent advances in shadow tomography and classical shadows have demonstrated that certain properties of quantum states, such as expectation values of low-weight observables, can be estimated efficiently using polynomially many measurements \cite{huang2020p}. This suggests that for structured encoders that produce states with low-weight Pauli representations (as discussed in the context of classically simulable QNNs), the distinguishability barrier may be overcome. Thus, while a quantum JEPA with perfectly Haar-random encoders is infeasible, a practical quantum JEPA with well-designed, structured encoders remains a direction worthy of investigation.

In general, it would be premature to claim that a quantum JEPA with barren plateaus is guaranteed to work. However, the convergence of three ideas, i.e., the exponential concentration of Haar-random states (ensuring anti-collapse), the geometric separation of quantum embeddings (preventing trivial solutions), and modern techniques for efficient quantum property estimation (mitigating distinguishability issues), makes this direction worthy of future investigation. The key idea underlying this direction is to explore the possibility of using barren plateaus as a resource rather than treating them as an obstacle. Other methodologies for achieving so would also be interesting.

\textbf{Pauli basis NNs.} A fundamental observation in quantum information is that the Pauli operators form a complete basis for the space of $2^n\times 2^n$ matrices. For $n$ qubits, the set of $n$-fold tensor products of single-qubit Pauli operators $\{I,X,Y,Z\}$
\begin{equation}
\mathcal{P}_n=\left\{\boldsymbol{P}_{\boldsymbol{\alpha}}=P_{\alpha_1}\otimes P_{\alpha_2}\otimes\cdots\otimes P_{\alpha_n}\mid P_{\alpha_i}\in\{I,X,Y,Z\}\right\},
\end{equation}
contains $4^n$ orthogonal basis elements satisfying $\operatorname{Tr}[P_{\boldsymbol{\alpha}} P_{\boldsymbol{\beta}}]=2^n\delta_{\boldsymbol{\alpha}\boldsymbol{\beta}}$. $\boldsymbol{\alpha}=(\alpha_1,\ldots,\alpha_n)$, with each components taking integers 0 to 3 (corresponding to one element in $\{I,X,Y,Z\}$). Consequently, any $2^n\times2^n$ matrix, whether the Hermitian ones (observable) or the unitary ones (quantum gate), or arbitrary linear operator, can be uniquely expanded as a linear combination
\begin{equation}
A=\sum_{\boldsymbol{\alpha}}c_{\boldsymbol{\alpha}}P_{\boldsymbol{\alpha}},~~c_{\boldsymbol{\alpha}}=\frac{1}{2^n}\operatorname{Tr}[P_{\boldsymbol{\alpha}} A]\in\mathbb{C}.
\end{equation}
This mathematical fact provides the foundation for a family of classical NN architectures that we refer to as \textit{Pauli basis neural networks} (PBNNs).

The core idea of a PBNN is to parameterize the model output directly as a linear combination of Pauli expectation values. Given an input $\mathbf{x}$ (classical or quantum), we first encode it into a quantum state $|\psi(\mathbf{x})\rangle$ via a fixed encoding circuit. The model then outputs
\begin{equation}\label{eq:PBNN}
f(\mathbf{x};\boldsymbol{\theta})=\sum_{\boldsymbol{\alpha}}\theta_{\boldsymbol{\alpha}}\langle \psi(\mathbf{x})|\boldsymbol{\boldsymbol{P_\alpha}}|\psi(\mathbf{x})\rangle,
\end{equation}
where $\{P_k\}$ is a selected subset of Pauli operators (typically those with low weight, i.e., acting nontrivially on few qubits), and $\boldsymbol{\theta}=(\theta_{\boldsymbol{\alpha}},\theta_{\boldsymbol{\beta}},\ldots,)\in\mathbb{R}^{4^n}$ are trainable classical coefficients. The quantum device serves as a fixed feature extractor producing Pauli expectation values, while all learning occurs in the classical combination coefficients.

An equivalent but more expressive variant allows the coefficients themselves to be parameterized by a classical NN
\begin{equation}
f(\mathbf{x};\boldsymbol{\phi},\boldsymbol{\psi})=\sum_{\boldsymbol{\alpha}} g_{\boldsymbol{\alpha}}(\mathbf{x};\boldsymbol{\phi})\,\langle\psi(\mathbf{x})|\boldsymbol{P_{\alpha}}|\psi(\mathbf{x})\rangle,
\end{equation}
where $g_{\boldsymbol{\alpha}}(\mathbf{x};\boldsymbol{\phi})$ is, for example, a small classical NN that maps the input to coefficient values. This hybrid architecture interpolates between purely classical models (when the quantum expectation values are trivial) and fully quantum models (when the coefficients are fixed and all learning happens in the quantum circuit). Several observations on the PBNNs can be given as follows. 

\begin{itemize}
    \item An intuitive observation is that PBNNs encompass several families of QNNs that have been proven classically simulable. For instance, the class of QNNs where the observable and circuit form a $2$-design can be shown to have output distributions that are efficiently approximated by truncated Pauli expansions \cite{McClean2018}. Similarly, the multi-scale entanglement renormalization ansatz (MERA) \cite{vidal2008}, a tensor network architecture for quantum many-body states, could be viewed as a specific PBNN with a hierarchical Pauli basis structure. The classical simulability of QCNNs discussed earlier \cite{Bermejo2026,angrisani2025} also follows from this observation. When the effective observable $U^\dagger(\boldsymbol{\theta}) OU(\boldsymbol{\theta})$ remains approximately in the low-weight Pauli subspace during training, the entire model can be simulated classically by tracking only polynomially many Pauli coefficients. Consequently, PBNNs could provide a unifying framework for the simulation of the above. 
    
    \item Besides, PBNNs could offer a practical framework for bridging the gap between low-expressivity classical models and high-expressivity quantum models. Low-expressivity models, such as linear models or shallow networks, are easy to train but lack the capacity to capture complex patterns. High-expressivity quantum models, such as deep parameterized quantum circuits, can represent intricate correlations but often suffer from barren plateaus and exponential training costs. The PBNN architecture naturally admits a spectrum of expressivity controlled by two factors: 1) the number $K$ of Pauli measurements included in the model, and 2) the richness of the coefficient $\theta_{\boldsymbol{\alpha}}$ (or the depth and complexity of network  $g_{\boldsymbol{\alpha}}(\mathbf{x};\boldsymbol{\phi})$) that combines these measurements. At one extreme, setting $K=O(1)$ and using fixed coefficients recovers a simple linear model in the Pauli expectation features, which is low expressivity but guaranteed trainability and no barren plateau issues. At the other extreme, taking $K$ exponential in the number of qubits $n$ and using deep classical coefficient networks approaches the expressivity of a fully parameterized quantum circuit, albeit with potentially exponential classical simulation cost. By smoothly interpolating between these extremes, PBNNs can serve as a \textit{workable connector} between regimes, allowing practitioners to trade off classical computational cost against quantum expressivity according to the requirements of their specific application.
    
    \item Moreover, the PBNN framework could also enable the construction of residual connections, which have been crucial for training deep classical NNs by mitigating the vanishing gradient problem and allowing information to flow directly through skip connections \cite{He2016ResNet}. It is known that implementing such residual connections in purely quantum settings remains challenging due to the no-cloning theorem, which forbids copying arbitrary quantum states. In a PBNN, which is basically classical NNs encoding the mathematical structures of quantum evolution, residual connections can be implemented directly at the level of Pauli coefficients rather than at the quantum state level. Specifically, one can define a residual block as follows
    \begin{equation}
    f^{(l+1)}(\mathbf{x})=f^{(l)}(\mathbf{x})+\sum_{\boldsymbol{\alpha}}\theta_{\boldsymbol{\alpha}}^{(l)}\langle \psi(\mathbf{x})|\boldsymbol{P_{{\alpha}}}^{(l)}|\psi(\mathbf{x})\rangle, 
    \end{equation}
    where $f^{(l)}$ is the output of the $l$-th residual block, $\{\boldsymbol{P_{\alpha}}^{(l)}\}$ are Pauli operators selected for the $l$-th block, $\theta_{\boldsymbol{\alpha}}^{(l)}$ are trainable coefficients, and $\langle\psi(\mathbf{x})|\boldsymbol{P_{\alpha}}^{(l)}|\psi(\mathbf{x})\rangle$ are quantum expectation values computed from the fixed encoding circuit. The skip connection, represented by the term $f^{(l)}(\mathbf{x})$ added to the new Pauli features, is purely a classical operation that does not require any quantum resources. One can also extend this idea to deeper architectures by stacking multiple residual blocks, where each block adds a new set of Pauli features to the accumulated output. 
\end{itemize}

Lastly, the function of a PBNN can be reorganized into a layerwise form under certain constraints. This can be seen more clearly through the following procedure. First, a reduction in computational steps can be achieved by a simple algebraic treatment. Suppose that $|\psi(\mathbf{x})\rangle$ in Eq.~(\ref{eq:PBNN}) is defined as $|\psi(\mathbf{x})\rangle=\mathcal{U}(\mathbf{x})|0\rangle^{\otimes n}$, obtained by encoding a classical input $\mathbf{x}$ into an $n$-qubit system via a fixed encoding circuit $\mathcal{U}(\mathbf{x})$. For a Pauli operator $P_{\boldsymbol{\alpha}}=P_{\alpha_1}\otimes P_{\alpha_2}\otimes \cdots\otimes P_{\alpha_n}$ (with each $P_{\alpha_i}\in\{I,X,Y,Z\}$), define the corresponding expectation value as
\begin{equation}
\langle P_{\boldsymbol{\alpha}}\rangle_{\mathbf{x}}=\langle\psi(\mathbf{x})|P_{\boldsymbol{\alpha}}|\psi(\mathbf{x})\rangle.
\end{equation}
If the encoding circuit $\mathcal{U}(\mathbf{x})$ is composed only of local operators, then the quantum state is a product state across qubits. In this case, the expectation value factorizes
\begin{equation}
\langle P_{\boldsymbol{\alpha}} \rangle_{\mathbf{x}} = \langle P_{\alpha_1} \rangle_{x_1} \langle P_{\alpha_2} \rangle_{x_2} \cdots \langle P_{\alpha_n} \rangle_{x_n},
\end{equation}
where $\langle P_{\alpha_i} \rangle_{x_i}$ denotes the expectation value of $P_{\alpha_i}$ on the $i$-th qubit state prepared by the local encoding. Such a setup is normal in the circuit-based quantum computation, where the entanglement usually emerges in the circuit part and the input is usually set to be a product state. 

The basic PBNN model then takes a linear combination of these expectation values with trainable classical coefficients $\boldsymbol{\theta}$
\begin{equation}
f(\mathbf{x};\boldsymbol{\theta})=\sum_{\boldsymbol{\alpha}} \theta_{\boldsymbol{\alpha}}\langle P_{\alpha_1}\rangle_{x_1}\langle P_{\alpha_2} \rangle_{x_2}\cdots\langle P_{\alpha_n}\rangle_{x_n}.
\end{equation}
By the distributive law, this expression can be reorganized into a nested form, shown by
\begin{equation}\label{Eq:distrPBNN}
f(\mathbf{x};\boldsymbol{\theta})=\tilde{\theta}_{0}\langle I\rangle_{x_1}\Bigl[ \tilde{\theta}_{00}\langle I\rangle_{x_2}(\cdots)+\tilde{\theta}_{01}\langle X \rangle_{x_2}(\cdots)+\tilde{\theta}_{02}\langle Y\rangle_{x_2}(\cdots)+\tilde{\theta}_{03}\langle Z\rangle_{x_2}(\cdots)\Bigr]+\tilde{\theta}_{1} \langle X\rangle_{x_1}\Bigl[\cdots\Bigr]+\cdots.
\end{equation} This nested reorganization reduces the computational complexity from $O(2^n\times 2^n)$ for direct summation over all $4^n$ Pauli strings to $O(n2^n)$ when evaluating the nested structure efficiently. This reduction is analogous to the improvement of the fast Fourier transform over the discrete Fourier transform, i.e., from $O(N^2)$ to $O(N\log N)$ operations for an input of size $N$, by exploiting a nested, divide-and-conquer structure \cite{cooley1965algorithm}. The connection between the two paradigms can be seen more clearly by setting $N=2^n$. However, note that this efficiency gain does not leads to a computation polynomial to qubit number $n$.

Second, consider a layerwise structure similar to Eq.~(\ref{Eq:distrPBNN}) but organized by qubits as layers. As the foundation, define a transformation function $\boldsymbol{W}_l$ that maps the feature vector of layer $l$ to layer $l+1$
\begin{equation}
\boldsymbol{W}_l(\boldsymbol{p}_{l+1},\boldsymbol{p}_l)=(\tilde{\boldsymbol{\theta}}'_{l+1}\circ\boldsymbol{p}_{l+1})\circ (\boldsymbol{M}_{l+1}\boldsymbol{p}_l),
\end{equation}
where $\boldsymbol{p}_l=(\langle I\rangle_{x_l},\langle X\rangle_{x_l},\langle Y\rangle_{x_l},\langle Z\rangle_{x_l})^\top\in\mathbb{R}^4$ is the vector of local expectation values for the $l$-th qubit. $\boldsymbol{M}_{l+1}\in\mathbb{C}^{4 \times 4}$ is a trainable matrix that linearly transforms $\boldsymbol{p}_l$. $\tilde{\boldsymbol{\theta}}'_{l+1}\in\mathbb{R}^4$ is a trainable coefficient vector, which is different from $\theta_{\boldsymbol{\alpha}}$ in values. $\circ$ denotes the Hadamard product \cite{sanjeet2025hadamard}. Then, the output of a $n$-layer PBNN can be expressed recursively as
\begin{equation}
f'(\mathbf{x};\tilde{\boldsymbol{\theta}}')=\boldsymbol{W}_K \left( \boldsymbol{p}_K, \boldsymbol{W}_{K-1} \left( \boldsymbol{p}_{K-1}, \cdots \boldsymbol{W}_2 \left( \boldsymbol{p}_2, \boldsymbol{W}_1 \left( \boldsymbol{p}_1, \tilde{\boldsymbol{\theta}}'_0 \circ \boldsymbol{p}_0 \right) \right) \cdots \right) \right),
\end{equation}
where $\boldsymbol{p}_0$ is an initial input vector.

By comparing the number of free parameters, one can observe that $f'(\mathbf{x};\tilde{\boldsymbol{\theta}}')$ is not generally equivalent to  $f(\mathbf{x};\boldsymbol{\theta})$ given by Eq. (\ref{Eq:distrPBNN}). $f(\mathbf{x};\boldsymbol{\theta})$ has up to $O(4^n)$ independent coefficients, while $f'(\mathbf{x};\tilde{\boldsymbol{\theta}}')$ has $O(n)$ parameters. Nevertheless, $f'(\mathbf{x};\tilde{\boldsymbol{\theta}}')$ is already highly expressive and can be computed efficiently. It can compute the cases involving special entanglements, though it is strictly less expressive than $f(\mathbf{x};\boldsymbol{\theta})$ due to its restricted parameterization.

\begin{figure}\centering
\includegraphics[width=.92\textwidth]{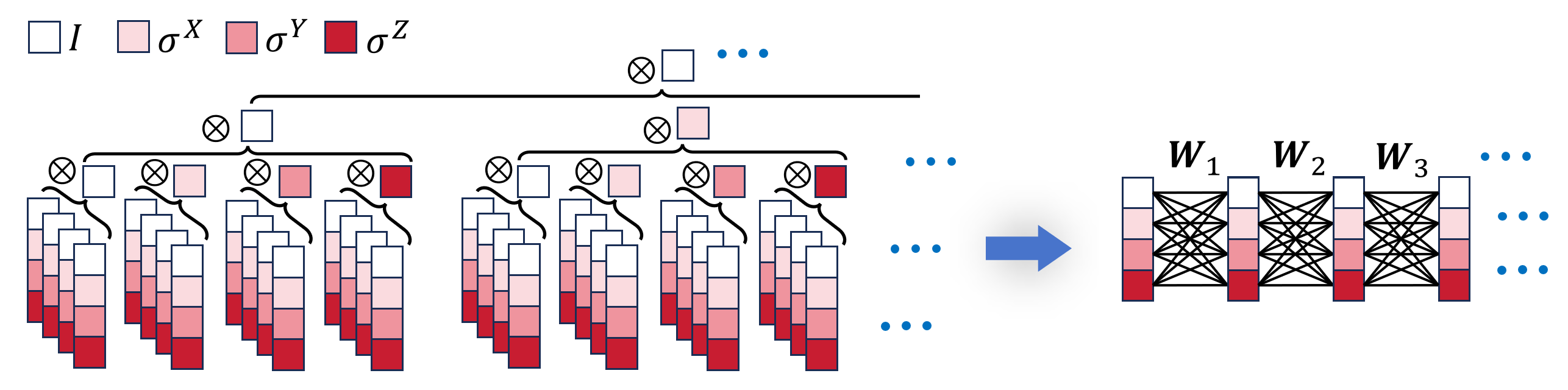}
\caption{An illustration of using decomposition as a network. The whole structure could be reformed as a layerwised structure under special constrains.}\label{FIG:8}
\end{figure}

Third, following the same line of reasoning, one can introduce nonlinear activation functions to increase the expressivity of $f'(\mathbf{x};\tilde{\boldsymbol{\theta}}')$, potentially making it a good approximation of $f(\mathbf{x};\boldsymbol{\theta})$ in a general sense. Let $\sigma$ denote a vectorized nonlinear activation function that acts elementwise on each component of a vector (possibly with different particular form per component). Then, a nonlinear PBNN can be defined as
\begin{equation}
\tilde{f}(\mathbf{x};\tilde{\boldsymbol{\theta}})=\sigma\left[\boldsymbol{W}_K \left(\boldsymbol{p}_K,\sigma\left[\boldsymbol{W}_{K-1}\left(\boldsymbol{p}_{K-1},\cdots\sigma\left[\boldsymbol{W}_2\left(\boldsymbol{p}_2,\sigma\left[\boldsymbol{W}_1\left(\boldsymbol{p}_1,\tilde{\boldsymbol{\theta}}'_0\circ\boldsymbol{p}_0\right)\right]\right)\right]\cdots\right)\right]\right)\right].
\end{equation}
A graphic illustration of this process is given in Fig.~\ref{FIG:8}. One significant difference between PBNNs and classical NNs lies in the mathematical operation connecting layers. In a classical NN, the basic operation is roughly additive. In contrast, the PBNN's layerwise operation is multiplicative, involving Hadamard products between transformed feature vectors. This multiplication arises naturally from the tensor product structure of Pauli operators and the factorization of expectation values under product inputs. Therefore, this primary architecture, which encodes quantum mathematical structures, may provide interesting properties not found in conventional NNs. Whether these properties are beneficial for machine learning remains an open question, which needs further exploration.

\vspace{2ex}
The field of QNNs and quantum machine learning represents a promising and rapidly evolving frontier in computational science. This manuscript has provided a comprehensive overview of the current state of QNNs, highlighting their diverse structures, applications, and potential advantages over classical NNs. We have explored various types of QNNs, including FCQNNs, QCNNs, EQNNs, QHNs, QBMs, and QRC. In addition to these specialized QNN architectures, we have also discussed the integration of QNNs into broader machine learning paradigms, such as QRL, QGL, and QTL. These composite approaches demonstrate the potential for QNNs to tackle a wide range of applications, from optimization and generative modeling to transfer learning and temporal pattern recognition. The exploration of hybrid quantum-classical models further underscores the versatility of QNNs, combining the strengths of both classical and quantum computing to achieve improved performance and efficiency. The manuscript has also highlighted the importance of optimization techniques in training QNNs, emphasizing the need for specialized algorithms that can navigate the unique challenges posed by quantum systems. Methods such as PSR, QNGD, and others have been introduced, showcasing the ongoing efforts to develop robust and efficient training strategies for QNNs.

Beyond the specific architectures and algorithms surveyed in this review, the most important inspiration emerging from QNNs and quantum machine learning at the current stage is twofold.

On the one hand, QNNs can be viewed as an inspiration for new classical models. Quantum mechanics offers a rich source of novel mathematical structures. The Pauli basis expansion, the tensor product structure of composite systems, the distributive law of linear operators, and the multiplicative layer interactions arising from quantum evolution all point toward unconventional algebraic frameworks that differ fundamentally from the additive, affine transformations underlying classical NNs. By examining these quantum mathematical structures in detail, zooming into the specific algebraic properties of the structures like Lie group and Lie algebra, researchers can identify whether such structures provide new benefits for classical machine learning tasks. Whether these structures translate into practical advantages remains an open question, but the exploration itself broadens the design space of machine learning models beyond the familiar additive paradigm. We notice several meaningful results have been reported along the line \cite{fan2024quantum,shi2024pretrained,yu2026quantum}.  

On the other hand, the development of QNNs and QML provides concrete hints for practitioners who wish to leverage currently available quantum platforms for machine learning. Near-term quantum devices, characterized by limited qubit counts (tens to hundreds), finite coherence times, and non-negligible noise, impose strict constraints on what can be effectively implemented. The survey of barren plateaus, for example, shows that deep circuits and global observables are likely to fail, while shallow circuits with local observables offer trainability. The classical simulability of certain QNN architectures, such as QCNNs, suggests that not every quantum-inspired model requires a quantum computer. Some can be efficiently simulated classically, and understanding this boundary is crucial for allocating resources. For those seeking to use current quantum platforms, the following directions are recommended based on the content above: (i) prioritize shallow circuits with local cost functions to avoid barren plateaus, (ii) consider hybrid quantum-classical models where the quantum device serves as a fixed feature extractor and classical networks handle the learning, (iii) leverage architectures with proven trainability, such as those with low-weight Pauli representations or those that avoid exponential concentration of gradients, and (iv) use optimization techniques such as the PSR for gradient estimation and QNGD for improved convergence, while recognizing that these methods do not fundamentally resolve barren plateaus.

In summary, the future of QNNs and quantum machine learning holds great promise, but the path forward requires a balanced perspective. Continued research into novel quantum architectures, optimization techniques, and hybrid quantum-classical models will deepen our understanding of quantum systems as learning machines and may eventually lead to genuine quantum advantage on fault-tolerant devices. Meanwhile, the most immediate impact may come from translating quantum-inspired mathematical structures, such as the Pauli basis expansion, the distributive law in tensor product spaces, and multiplicative layer interactions, into new classical models that can be deployed on conventional hardware. This dual approach ensures that progress in the field is not solely dependent on the maturation of quantum hardware, but also enriches classical machine learning with fresh design principles. The ongoing exploration of novel QNN architectures, optimization techniques, and hybrid quantum-classical models will be crucial in unlocking the full potential of quantum machine learning and driving the next wave of innovation in computational intelligence. Whether the ultimate contribution of quantum machine learning to the broader field of machine learning turns out to be new quantum algorithms, new classical algorithms inspired by quantum mathematics, or both, the journey of discovery promises to be as valuable as the destination.

\section*{Acknowledgement} 
This work is supported by the National Key R \& D Program of China under Grant No. 2022YFA1404904 and the National Natural Science Foundation of China (No. 12234004 and No. 12474355).









\printcredits

\bibliographystyle{model1-num-names}
\bibliography{refs}



\end{document}